\documentclass[longauth]{aa}

\usepackage{graphicx}
\usepackage{txfonts}
\usepackage{amsmath}
\usepackage{booktabs}
\usepackage{hyperref}

\begin{document}

   \title{Dust production in the harsh environment of Sgr~A*}
   \subtitle{MIRI/JWST observation of the O-rich AGB star IRS~3}
   \titlerunning{}

	\author{F. Pei{\ss}ker\inst{\ref{inst1},\ref{inst2}}
        \and M. Garc\'{\i}a Mar\'{\i}n\inst{\ref{inst3}}
        \and A. Eckart\inst{\ref{inst1},\ref{inst4}}
        \and G. Wright\inst{\ref{inst5}}
        \and O. C. Jones\inst{\ref{inst5}}
        \and D. Dicken\inst{\ref{inst5}}
        \and A. Alonso Herrero\inst{\ref{inst6}}
        \and D. Rouan\inst{\ref{inst7}}
        \and D. Law\inst{\ref{inst8}}
        \and T. Böker\inst{\ref{inst3}}
        \and T. Henning\inst{\ref{inst9}}
        \and M. Baes\inst{\ref{inst10}}
        \and A. Labiano\inst{\ref{inst11}}
        \and L. Pantoni\inst{\ref{inst10}}
        \and L. Hermosa Mu{\~n}oz\inst{\ref{inst6}}
        \and P.O. Lagage\inst{\ref{inst13}}
        \and P. van der Werf\inst{\ref{inst15}}
        \and G. {\"O}stlin\inst{\ref{inst16}}        
        \and J. A. D. L. Blommaert\inst{\ref{inst17}}
        \and M. G{\"u}del\inst{\ref{inst18},\ref{inst19}}
        \and P. Guillard\inst{\ref{inst20}}
        } 

	\institute{I.Physikalisches Institut der Universit\"at zu K\"oln, Z\"ulpicher Str. 77, 50937 K\"oln, Germany\label{inst1} \and
	\email{peissker@ph1.uni-koeln.de}\label{inst2}
    \and European Space Agency (ESA), ESA Office, Space Telescope Science Institute, 3700 San Martin Drive, Baltimore, MD 21218, USA\label{inst3}
    \and Max-Planck-Institut f\"ur Radioastronomie, Auf dem H\"ugel 69, 53121 Bonn, Germany\label{inst4}
    \and UK Astronomy Technology Centre, Royal Observatory, Blackford Hill, Edinburgh EH9 3HJ, UK\label{inst5}
    \and Centro de Astrobiología (CAB), CSIC-INTA, Camino Bajo del Castillo s/n, E-28692 Villanueva de la Cañada, Madrid, Spain\label{inst6}
    \and LIRA, Observatoire de Paris, Université PSL, Sorbonne Université, Université Paris Cité, CY Cergy Paris Université, CNRS, 92190 Meudon, France\label{inst7}
    \and Space Telescope Science Institute, 3700 San Martin Drive, Baltimore, MD 21218, USA\label{inst8}
    \and Max-Planck-Institut für Astronomie, Königstuhl 17, 69120 Heidelberg, Germany\label{inst9}
    \and Department of Physics and Astronomy, Universiteit Gent, Proeftuinstraat 86 N3, B-9000 Ghent, Belgium\label{inst10}
    \and Telespazio UK SL for ESA, ESAC, Camino Bajo del Castillo s/n, E-28692 Villanueva de la Cañada, Madrid, Spain\label{inst11}    
    \and Université Paris Cité, Université Paris-Saclay, CEA, CNRS, AIM, F-91191 Gif-sur-Yvette, France\label{inst13}
    \and Leiden Observatory, Leiden University, PO Box 9513, 2300 RA Leiden, The Netherlands\label{inst15}  
    \and Department of Astronomy, Oskar Klein Centre, Stockholm University, AlbaNova University Center, 10691 Stockholm, Sweden\label{inst16}   
    \and Astronomy and Astrophysics Research Group, Department of Physics and Astrophysics, Vrije Universiteit Brussel, Pleinlaan 2, B-1050 Brussels, Belgium\label{inst17}
    \and Dept. of Astrophysics, University of Vienna, T\"urkenschanzstr. 17, A-1180 Vienna, Austria\label{inst18}
    \and ETH Z\"urich, Institute for Particle Physics and Astrophysics, Wolfgang-Pauli-Str. 27, 8093 Z\"urich, Switzerland\label{inst19}
    \and Sorbonne Université, CNRS, Institut d’Astrophysique de Paris, 98 bis bd Arago, 75014 Paris, France\label{inst20}}

   \date{Received XXX; accepted XXX}

  \abstract
   {Studies of the interstellar medium (ISM) have frequently revealed signatures of the dust produced in the envelopes of Asymptotic Giant Branch (AGB) stars, demonstrating a connection between the dust composition of the ISM and that of AGB stellar envelopes. Investigating this relationship in the extreme, radiation-dominated environment surrounding Sgr~A*, the center of our own galaxy, reveals how such conditions might influence dust composition and the recycling of material in galactic centers.}
   {IRS~3, the brightest L band source in the Galactic Center and most prominent AGB star within the {inner parsec} of the Milky Way, is embedded in a dusty envelope with an estimated radius of $\sim$10000 AU. We aim to conduct a comprehensive spectral analysis to more tightly constrain the dust composition and line‑emitting species within the envelope of IRS~3 in the immediate vicinity of Sgr~A*.}
   {As part of the guaranteed time observations (GTO) program Mid-Infrared Characterisation of Nearby Iconic galaxy Centres (MICONIC) performed with the Mid-Infrared Instrument (MIRI) on board the James Webb Space Telescope (JWST), we observed the {inner parsec} of the Milky Way in 2025. We used the MIRI Medium Resolution Spectrometer (MRS) to study the spectroscopic characteristics of the AGB star IRS~3 located about 0.17 parsecs in projection away from Sgr~A*.}
   {After correcting for foreground dust extinction along the line of sight using a stellar‑based mid‑infrared extinction curve, we find a broad 9.7 $\mu$m silicate stretching feature accompanied by an O-Si-O bending mode absorption feature at 18.5 $\mu$m in the spectrum of IRS~3. Using the radiative transfer code Hyperion, we model the observed spectrum of IRS~3 {with a best-fit stellar luminosity of 60000 L$_{\odot}$. Based on our model, we find a multi-shell configuration that shows} a significant temperature gradient between the inner and outer layers of the envelope. A combination of alumina and amorphous silicates accurately reproduces the observed spectrum. {We infer a stellar mass of about 6 M$_{\odot}$ and a corresponding age of $\approx\,72\,$Myr.} For the first time, we find clear signs of water in the envelope of IRS~3.}
   {Based on our MIRI MRS observations and the modeled spectrum, we conclude that the dusty envelope of IRS~3 may be characterized by a shell-like distribution with a temperature gradient of $\approx$ 1000 K. Analysis of the MIRI MRS spectrum reveals the presence of amorphous silicates, Al$_2$O$_3$, and H$_2$O. A ratio of $3.5\pm0.1$ between the optical depth of the silicate Si-O stretching feature at 9.7 $\mu$m and the O-Si-O bending mode classifies IRS~3 as an O-rich star. Furthermore, {assuming a wind velocity of v$\rm _w\,=\,$15 km/s, we} estimate a mass-loss rate of $\rm 6 \times 10^{-5} \, M_{\odot}\,{\rm yr^{-1}}$ and {identify} H$_2$O in the envelope of IRS~3. This finding demonstrates that the harsh radiation-dominated environment of Sgr~A* does not inhibit dust formation or the survival of molecular species such as H$_2$O.}

   \keywords{Galaxy: center, Stars: AGB and post-AGB, Stars: evolution, Stars: chemically peculiar, Infrared: stars, Infrared: ISM, ISM: dust, extinction, Stars: black holes}

   \maketitle

\section{Introduction} 
\label{sec:intro}

Once a low- and intermediate-mass star ($\rm \leq\,8\,M_{\odot}$) exhausts the helium supply in its core, it enters the Asymptotic Giant Branch (AGB) phase \citep{Iben1983, Vassiliadis1993, Busso1999, Herwig2005}. The AGB phase of a star is accompanied by pulsations that eject gas far away from the star, where it cools and condenses, forming dusty circumstellar envelopes \citep{Hoefner1997}. Due to the large size of these envelopes, a few $10^3-10^4$ AU, interactions with the Interstellar Medium (ISM) are inevitable \citep{Zajacek2020,Maercker2022}. Through their substantial mass loss, AGB stars are significant sources of dust injected into the interstellar medium \citep[e.g.,][and references therein]{Ferrarotti2006, Hoefner2018}. In general, the inner structure of an AGB star can be classified in three groups: {carbon-rich (C-type) stars showing a C/O ratio $>1$, M-type (oxygen-rich) stars exhibiting a C/O ratio $<1$,} and S-type AGB stars with a C/O ratio of $\approx1$. Most of these late-type stars are embedded in an envelope whose composition is influenced by the AGB stellar type. 

Interestingly, some of the envelopes of AGB stars have been shown to indicate the presence of a binary \citep{Homan2020, Decin2020}. Although the ISM or putative companions are local sources of interaction, \cite{Zajacek2020} and \cite{Kurfuerst2025} have proposed that jets that originate from supermassive black holes are responsible for stripped-away envelopes of AGB stars in the Galactic Center. Although a potential jet-envelope interaction may explain the missing red giants, the authors of \cite{Zajacek2020} focus mainly on distances of 0.04 pc from Sgr~A*. In contrast, the prominent and bright mid-infrared star and bow shock source IRS~3 is located a projected distance of about 0.17 pc from Sgr~A* \citep{Chiar2000, Kemper2004, Pott2008}. Due to this substantially larger distance of IRS~3 from Sgr~A*, the dominant mechanism for producing the bow shock of the AGB star is the interaction with the ISM \citep{Wilkin1996, Wilkin2000, Christie2016}.

However, the most striking feature of IRS~3 is its envelope that roughly points towards the position of Sgr~A*, leading to speculations about the interaction between the envelope and an undetected jet that originates at the position of the SMBH \citep{Eckart2006} and could explain the low number of detected AGB stars in the {inner parsec} \citep{Schoedel2009, Gallego-Cano2018}. While a jet-star interaction or the presence of a binary related to IRS~3 may be revealed in future observation campaigns, the high sensitivity and broad wavelength range of our MIRI data allows us to investigate other important questions: Does the classification of a C-rich AGB star, proposed by \cite{Pott2008}, hold? What is the nature of the dust produced in the envelope of IRS~3? Does the AGB star produce fresh dust or will the envelope be stripped away in the close future \citep{Zajacek2020}? What is the related mass loss rate of the AGB star? 

Considering the envelope of IRS~3 with a radius of $\sim$10000 AU, another pressing question addresses the uniqueness of the star. Given its approximate age {of a few Gyr \citep{Pfuhl2011, Gallego-Cano2026}}, the Nuclear Stellar Cluster, and thus the {inner parsec} with its millions of stars, should contain more observable AGB sources with prominent envelopes \citep{Gallego-Cano2018}. Yet, IRS~3 dominates the L- (3.6 $\mu$m) and M-band (4.5 $\mu$m) observations of the vicinity of Sgr~A* and seems to be the only representative AGB star with a large dusty envelope. Like the ``paradox of youth'' that describes the surprisingly young age of the stars close to Sgr A* \citep{Ghez2003}, the paradox of the missing red giants marks another debated question \citep{Zajacek2020}. In light of these {paradoxa extrema}, understanding the basic parameters of IRS3 gains heightened importance. 

In this work, we analyze the observations of the AGB star IRS3 carried out with the MIRI MRS \citep{Wright2015, Wright2023}, on board the James Webb Space Telescope \citep{Gardner2023}. For the first time, we show an extinction corrected mid-infrared spectrum of IRS~3 between 4.9 $\mu$m and 27.9 $\mu$m. We apply established Galactic Center extinction laws to strengthen the robustness of the analysis. 
The presented analysis is structured as follows: In Sec. \ref{sec:data}, we describe the observations, data reduction and analysis process. The results of the MIRI/JWST observations are shown in Sec. \ref{sec:results}. Finally, we discuss our results in Sec. \ref{sec:discuss_conclusion} followed by the conclusion in Sec. \ref{sec:conclusion}.

\section{Data and Analysis} 
\label{sec:data}

This section describes the observations and methods used in this work. We outline the extinction correction and the setups used for the models constructed with Hyperion. Furthermore, we indicate the steps done for the synthetic spectrum based on HITRAN transition line lists to identify potential molecular species in the spectrum of IRS~3. 

\subsection{Observation and data reduction}

IRS~3 was observed with the MIRI Medium Resolution Spectrometer (MRS; \citealt{Wells2015,Argyriou2023}) onboard JWST on 21st April, 2025 as part of the guaranteed time observations (GTO) Mid-Infrared Characterisation of Nearby Iconic galaxy Centres (MICONIC) program 1266 (PI: M. Garcia Marin).
\begin{figure}[htbp!]
	\centering
	\includegraphics[width=.5\textwidth]{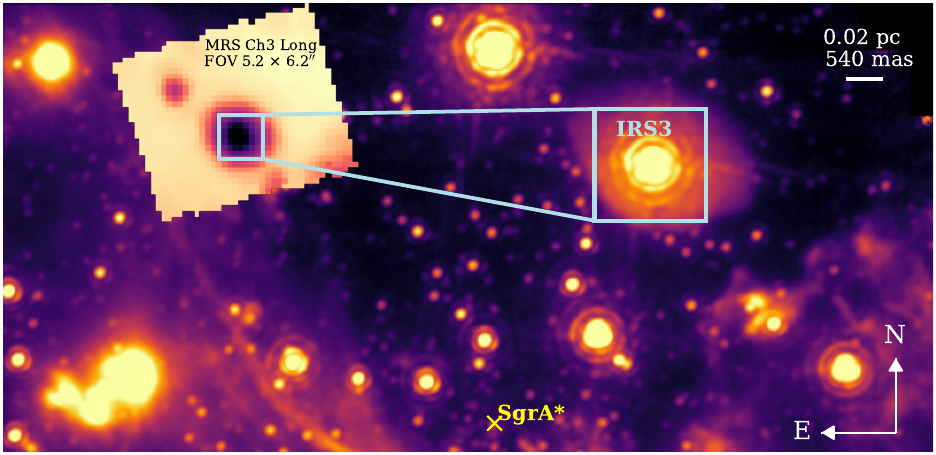}
	\caption{Mid-infrared image of the IRS~3 environment, observed with NACO (VLT) and MIRI/MRS (JWST). The background image is observed in the L-band (3.6 $\mu$m), the yellow inset shows a slice from the Channel 3 Long data cube, which covers (15.41–17.98 $\mu$m). The light blue boxes highlight an area of 1.6"$\times$1.6". The SMBH Sgr~A* is marked with a yellow $\times$, located at R.A. 17h45m40.05s and DEC -29:00:28.120. The projected distance between Sgr~A* and IRS~3 is about 4.4 arcsec which equals 0.17 parsec. The NACO FOV is about 13.5"$\times$6.5". In the NACO and MIRI MRS image, the emission of IRS~3 is dominated by the extended envelope of the AGB star.}
\label{fig:finding_chart}
\end{figure}
The observations used all MRS bands, delivering a total wavelength coverage from 4.9 to 27.9 $\mu$m over a FOV ranging from 3.2$\,\times\,$3.7 arcsec to 6.6$\,\times\,$7.7 arcsec.
We implemented a standard 4$-$point dither pattern, with three integrations of 15 groups in the FASTR1 readout pattern obtained at each dither position, for a total on$-$source time of 1500~s.
A background observation was also obtained, using a 2$-$dither pattern and the same detector set$-$up as the science observation. We note that MIRI simultaneous imaging was also taken, but that data are not included in this letter. The MRS data were processed with version 1.20.2 of the JWST calibration pipeline \citep{Bushouse2025}, context jwst\_1464.pmap. 

We use the standard extraction of the MIRI pipeline with a radius twice as large as the PSF to match the extent of the envelope of IRS~3. The MIRI MRS PSF FWHM ranges from about $0.39"$ to almost $1.0"$ at 25 $\mu m$; to account for this change the calibration pipeline uses a tapered conical aperture that grows with wavelength. 

The data was affected by strong saturation in several bands; since this saturation occurred in fewer than 2 groups the standard pipeline was unable to derive a slope, resulting in severe artifacts at the affected wavelengths. To remediate this, we applied a custom reduction for single-group data that is described in the Appendix \ref{appendix:post_processing}, along with additional processing details such as addressing the partial spectral discontinuities due to overlapping channels. 

\subsection{Extinction correction}

\begin{figure}[htbp!]
	\centering
	\includegraphics[width=.5\textwidth]{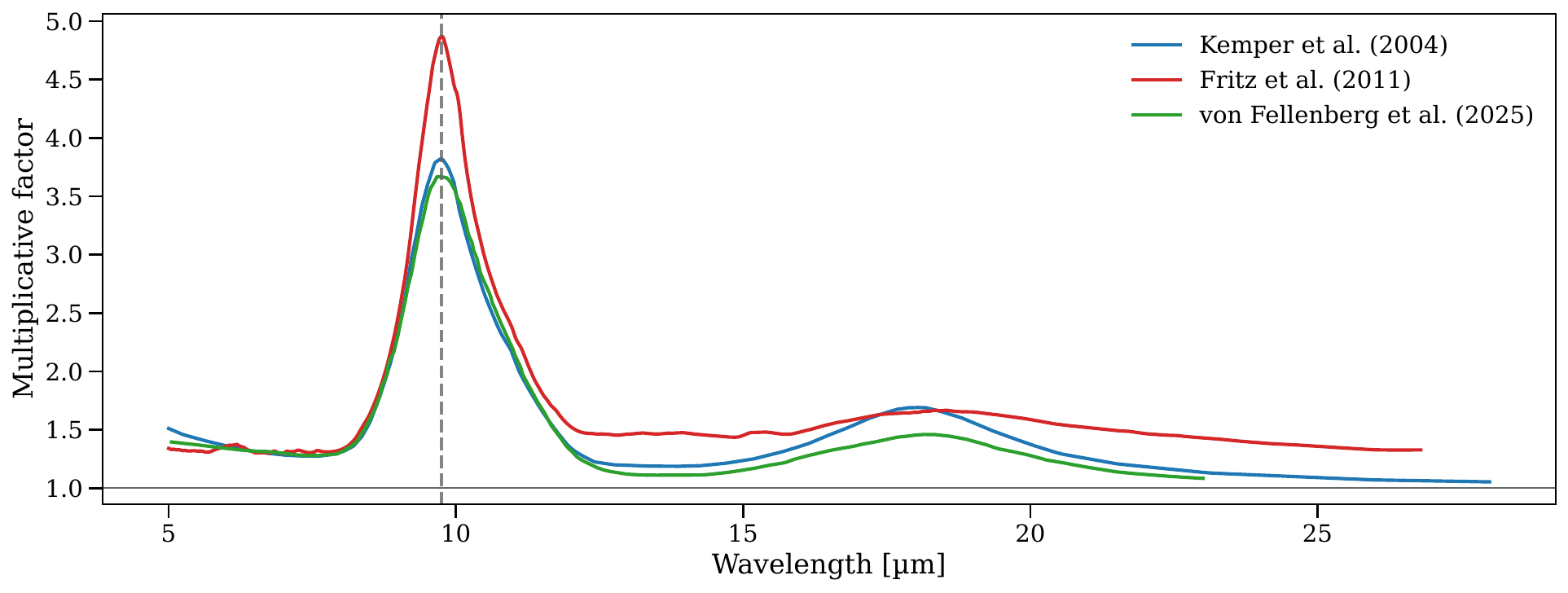}
	\caption{Comparison of available extinction laws for the {inner parsec} of the Galactic Center. Here, we plot the factor $\rm 10^{0.4\cdot A_{\rm \lambda}}$ against the wavelength. The observed flux is multiplied by this factor to achieve the dereddened spectrum (see text for details).}
\label{fig:multiplicative_factor_comp}
\end{figure}

The spectral analysis of a deeply dust-embedded source is sensitive to the extinction correction. Due to this, we will compare all available extinction laws for the {inner parsec}. However, while the extinction laws derived by \cite{Fritz2011} and \cite{Fellenberg2025} provide valuable tools for the {inner parsec} based on transition lines or continuum models, we prefer an empirically calibrated and stellar-motivated approach\footnote{The authors of \cite{Kemper2004} compare large beam Infrared-Space-Observatory (ISO) observations dominated by the Wolf-Rayet star IRS 1W \citep{Paumard2006} with the Quintuplet-cluster sources and reconstruct the extinction law used in this study.}. Therefore, we apply the foreground extinction correction to the IRS~3 spectrum using the normalized MIR stellar-based extinction curve of \cite{Kemper2004}. The dereddened flux F$_{\rm der}$($\rm \lambda$) of IRS~3 is described by
\begin{equation}
   \rm F_{\rm der}(\lambda)\,=\,F_{\rm obs}(\lambda)\,\times\,10^{0.4\cdot A_{\rm \lambda}}
\end{equation}
where F$_{\rm obs}$($\lambda$) is the observed flux (i.e. intrinsic and foreground) and A$_{\rm \lambda}$ is the wavelength-dependent extinction. Due to the normalization of the extinction law from Kemper et al., we set A$_{\rm \lambda}$= A$_{\rm fg}$k($\lambda$). Here, k($\lambda$) represents the normalized extinction law and A$_{\rm fg}$ a foreground factor that we set to 0.3 mag. At $\rm \lambda\,=$ 9.7 $\mu$m with
\begin{equation}
    \rm A(\rm \lambda)\,=\,A_{\rm fg}k(\rm \lambda) = \rm0.3\,mag\times 4.8\,=\,1.44\,mag
\end{equation}
this translates into a multiplicative factor of $\rm 10^{0.4\times1.44}\,=\,3.66$, consistent with the continuum model derived by \cite{Fellenberg2025}. For qualitative comparison, we inspect the multiplicative factor for all three available extinction laws as displayed in Fig. \ref{fig:multiplicative_factor_comp}.
The major difference between the three mid-infrared extinction laws is around 9.7 $\mu$m, explained by a different optical depth of the used objects/regions to construct the models. Regardless of the difference in optical depth, the width of the 9.7 $\mu$m silicate feature is comparable for all three models.

As evident in Fig. \ref{fig:multiplicative_factor_comp}, the spectral region between 5.0-7.7 $\mu$m shows an absence of features with an almost flat shape. Consequently, we scale the extinction law from \cite{Kemper2004} to the continuum-dominated emission of IRS~3 in the spectral range of $\rm 5.0-7.7\,\mu m$. In other words, we use a continuum-dominated spectral region that is not impacted by silicate absorption features and align the extinction law with the reddened spectrum of IRS~3. As mentioned, we set A$_{\rm fg}\,=\,0.3$ to achieve multiplicative factors comparable to \cite{Fritz2011} and \cite{Fellenberg2025}. We note that higher values of A$_{\rm fg}$ yield inconsistent extinction levels at 9.7 $\mu$m.

\subsection{Hyperion}

To analyze the composition of IRS~3, we use the 3D Monte-Carlo radiation transfer code Hyperion \citep{Robitaille2011}. In total, we tested 10$^5$ models with different parameter settings that are explained in the following. 
\begin{table*}[t!]
\centering
\caption{Input parameters for Hyperion to model the observed spectrum of IRS~3.}
\label{tab:irs3_parameter_ranges}
\begin{tabular}{lcccc}
\hline
Parameter 
& Inner region
& Shell 1 
& Shell 2 
& Shell 3 \\
\hline
Temperature $T$ [K]
 & 1200
 & 280--300
 & 180
 & 80--100 \\

Characteristic radius [AU]
 & $\sim$63
 & $\sim$949
 & $\sim$2214
 & $\sim$6325 \\

Density exponent $p$
 & 2.0 (in), 1.2 (out)
 & 0.1--0.2
 & 2.0
 & 2.5--3.0 \\

Density $\rho_0$ [g\,cm$^{-3}$]
 & $4\times10^{-15}$ (in), $3\times10^{-19}$ (out)
 & $1\times10^{-20}$
 & $3-10\times10^{-20}$
 & $3-10\times10^{-22}$ \\

Dust species
 & Alumina $\&$ Silicates
 & Silicates
 & Silicates
 & Silicates \\

\hline
\end{tabular}
\tablefoot{The characteristic radius is calculated from the stellar luminosity and the indicated temperature T. Hyperion does not require a specific location of the included shells but rather the approximate radius of the system, which is measured to be around R$_{\rm total}\approx 10^4$AU. Please see the text for details.}
\end{table*}

To explore the parameter space of our model, we use different setups to reconstruct the dereddened spectrum of IRS~3. Since multiple shell structures are associated with AGB stars \citep{Zhao-Geisler2012, Adam2019, Decin2020, Randall2020}, we adapt this envelope composition for our models.

For the initial setup of the central blackbody, we explore a temperature range between 1500-3000 K. We set the final stellar temperature in agreement with \cite{Pott2008} to 2800 K. For the stellar luminosity, we have inspected a range between {6000-60000 L$_{\odot}$. This range is motivated by observations of O-rich AGB stars in the galactic bulge \citep{Olofsson2022} but also the 50000 L$_{\odot}$ as found by \cite{Pott2008}. The final best fit characteristic luminosity is set to $\rm L\,=\,60000\,L_{\odot}$ based on the Root Mean Squared Error (RMSE) evaluation.}
For the dust, we adapt the silicates of O-rich late type stars from \cite{Ossenkopf1992} and alumina (Al$_2$O$_3$) from \cite{Begemann1997} and \cite{Suh2016}. Based on the L band emission of IRS~3 shown in Fig. \ref{fig:finding_chart}, we measure a projected envelope size of 0.1 parsec in diameter. This translates into an envelope radius of R$_{\rm total}\approx 10^4$ AU. Potential colder dust that may contribute to the envelope is below the detection limit or is photoevaporated due to the UV radiation in the Nuclear Stellar Cluster. Internally, the envelope is treated as a continuous component by Hyperion. However, we can define substructures that can be distinguished by their density, temperature, centrosymmetric radius, and dust composition. For the envelope of IRS~3, these substructures are individual shells located at increasing characteristic radii, defined as
\begin{equation}
\label{eq:sub}
    \rm r_{in,i}(T_i)\,=\,\bigg(\frac{L_\star}{16\pi k_{B}T_i^4} \bigg)^{\frac{1}{2}}
\end{equation}
where $\rm L_\star$ is the stellar luminosity, T the temperature, i the index of the related shell, and k$\rm_{B}$ the Boltzmann constant. Further, the density $\rm \rho$ is implemented as
\begin{equation}
    \rm \rho_i(r)\,=\,\rho_{0,i}\bigg(\frac{r}{r_{in,i}}\bigg)^{-p_i}
\end{equation}
where p is the density exponent, $\rm \rho_i$ the density of each shell, and r is the radius related to $\rm r_{in,i}\,\leq\, r_{out,i}$. It is important to note that the extent of these shells does not have sharp edges but transition zones interpolated by Hyperion. This becomes important for the inner regions due to the sublimation radius impacted by the temperature of the modeled blackbody.

Naturally, the inner and warmer regions with temperatures of about 1200 K favor the presence of alumina \citep{Begemann1997}, while the outer and colder regions can be modeled with amorphous silicates according to \cite{Ossenkopf1992}. The final parameter grid is listed in Table \ref{tab:irs3_parameter_ranges}, where we define an inner region in addition to the three shells. This inner region is {potentially} populated with Al$_2$O$_3$ in terms of the sublimation radius R$_{\rm sub}$ using Eq. \ref{eq:sub} with a sublimation temperature of T$_{\rm sub}$ = 1200 K.

In our preferred model, the density of the inner part of this region follows an exponent of p = 2.0. The outer part of the inner region is populated with silicates and follows a density exponent of p = 1.2, representing a transition zone between the inner 20 AU and the surrounding shells (Table \ref{tab:irs3_parameter_ranges}).

\subsection{HITRAN}

Since the IRS~3 spectrum shows a prominent absorption band together with a broader absorption feature between 6.12 and 6.25 $\mu$m (see Fig. \ref{fig:IRS3_spectrum}), we construct a synthetic spectrum using HITRAN\footnote{\url{https://hitran.org/}} to model the molecular species in the indicated wavelength range \citep{Gordon2026}.
In agreement with \cite{Sloan2015}, we associate the narrow absorption structure in this wavelength range primarily with H$_2$O, while the broader component remains ambiguous, as also discussed by \cite{Keane2001}. For a qualitative analysis, we used H$_2$O line transition lists from the HITRAN database.

The H$_2$O transition was modeled with a Voigt profile that includes thermal, turbulent, and macroscopic broadening, as well as pressure broadening and pressure-induced line shifts. The model includes an H$_2$O component with a temperature of 700 K and column densities of $1.5 \times 10^{17}$ cm$^{-2}$. The total optical depth $\rm \tau(\lambda)$ is defined as
\begin{equation}
    \rm \tau(\lambda)\,=\,\tau_{mol}(\lambda)\,+\,\tau_{G}(\lambda)
\end{equation}
where indicates $\rm \tau_{mol}(\lambda)$ the molecular optical depth and $\rm \tau_{G}(\lambda)$ a residual Gaussian component. The molecular optical depth is defined by
\begin{equation}
    \rm \tau_{mol}(\lambda)\,=\,\sum_i\,N_i S_i(T)\phi_i(\lambda)
\end{equation}
where $\rm N_i$ defines the column density, $\rm S_i$ the line strength, and $\phi_i$ the Voigt profile of the related species. {For the Gaussian component of the Voigt profile, we include non-thermal velocity dispersions that are described with v$_{\rm turb}$ and v$_{\rm mac}$, respectively. The turbulence is represented by v$_{\rm turb}$ whereas the macroscopic velocity term is described by v$_{\rm mac}$. In addition, the Lorentz component of the Voigt profile is modeled with a pressure p of 1.0$^{-6}$ atm.} The settings used for the synthetic spectrum are listed in Table \ref{tab:hitran}. For the residual Gauss components, we use
\begin{equation}
    \rm \tau_{G}(\lambda)\,=\,\sum_j\,\tau_{0,j}exp\bigg[- \frac{(\lambda-\lambda_{0,j})^2}{2(FWHM)_j^2}\bigg]
\end{equation}
where $\rm \tau_{0,j}$ is the optical depth, FWHM the full-width-half-maximum, and $\rm \lambda_{0,j}$ the wavelength of the individual transition lines (Table \ref{tab:gaussian_components}). Consequently, we fit the underlying continuum in the wavelength range 6.0-6.25 $\mu$m and get F$_{\rm cont}$ with
\begin{equation}
    \rm F_{\rm syn}(\lambda)\,=\,F_{\rm cont}(\lambda)exp(\tau(\lambda))
\end{equation}
as the flux of the synthetic spectrum as a function of wavelength $\lambda$.
\begin{table}[h!]
	\centering
    \caption{Basic parameters to model the synthetic spectrum based on HITRAN transition lists.}
	\begin{tabular}{|cc|}
		\hline \hline
		   Parameter & H$_2$O  \\ \hline
          Temperature [K]    &  700  \\
            Column density  [cm$^{-2}$]    &  1.5$\times$10$^{17}$  \\ 
            Pressure [atm]     &  1.0$^{-6}$ \\
          v$_{\rm mac}$ [km/s] &  4.0  \\ 
          v$_{\rm tur}$ [km/s] &  1.0  \\ \hline
	\end{tabular}
	\tablefoot{In addition {to H$_2$O}, we identify features in the spectrum that are approximated by a Gaussian fit model.}
	\label{tab:hitran}
\end{table}
The resulting transmission spectrum was convolved to the MIRI spectral resolution of $R = 3000$. 

In addition to the transition line residual mentioned above, we identify two broad absorption features between 6.05-6.10 $\mu$m and 6.12-6.25 $\mu$m. To approximate this mismatch, we added Gaussian absorption components to the synthetic spectrum F$_{\rm syn}$. The best-fit model was selected by minimizing the RMSE in comparison with the dereddened spectrum F$_{\rm der}(\lambda)$ in this wavelength range. 

\section{Results}
\label{sec:results}

In this section, we present the results of the analysis of the MIRI MRS observations. We show the dereddened spectrum, the modeled SED using Hyperion, and the analysis of a spectral range with prominent absorption lines.
\subsection{Observed spectrum of IRS~3}

Based on the MIRI MRS observations, we present the results of the data analysis outlined in Sec. \ref{sec:data}. 
\begin{figure*}[h!]
	\centering
    \sidecaption
	\includegraphics[width=12cm]{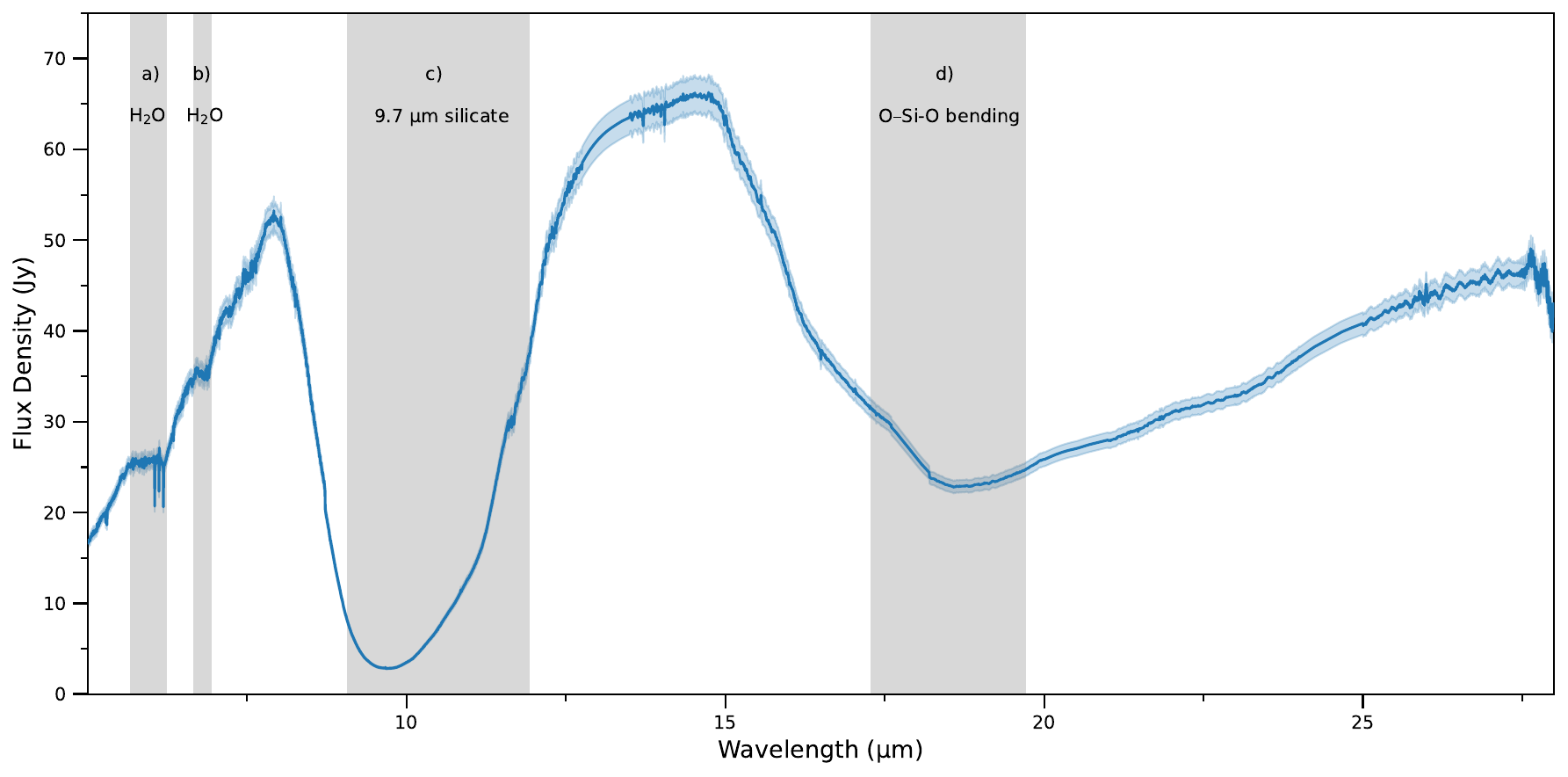}
	\caption{Dereddened MIRI spectrum of IRS~3 observed in 2025. We identify four spectral regions that are marked with gray bars, indicating features related to the nature of the AGB star. {These four spectral regions are described with a), b), c), and d). They cover a spectral range of 5.67-6.22 $\rm \mu$m, 6.66-6.95 $\rm \mu$m, 9.07-11.91 $\rm \mu$m, and 17.28-19.71 $\rm \mu$m.} The two most prominent regions include the $\rm 9.7\,\mu m$ silicate Si-O stretching feature and the O-Si-O bending mode at $\rm 18.5\,\mu m$.  Between 6-8 $\mu$m, we find spectral signatures of water absorption. The blue-shaded region of the spectrum indicates a 1$\sigma$ uncertainty of 3$\%$ of the foreground-corrected flux of IRS~3.}
\label{fig:IRS3_spectrum}
\end{figure*}
The extinction corrected spectrum of IRS~3 presented in Fig. \ref{fig:IRS3_spectrum} is extracted from all four MIRI channels (consider Fig. \ref{appendix:raw_spectrum} for a comparison to the raw spectrum). Please consider Sec. \ref{sec:data} for a detailed description of the construction of the spectrum.

In the dereddened spectrum of IRS~3, we detect forbidden emission lines that are related to the foreground. Since our extinction correction is explicitly based on the continuum, neglecting molecular lines, we individually correct emission lines associated with the foreground. We identify two prominent emission lines [SIII] at 18.7 $\mu$m reported by \cite{Herter1983} and [FeIII] at 22.9 $\mu$m \citep{Lutz1993, Peissker2020b} by interpolating the continuum.

In the dereddened spectrum presented in Fig. \ref{fig:IRS3_spectrum}, we detect a prominent Si-O silicate absorption feature at $\rm 9.7\,\mu m$ with an accompanying O-Si-O bending mode at about $\rm 18.5\,\mu m$ after we applied the extinction correction. Both features are marked with gray bars in Fig. \ref{fig:IRS3_spectrum}. In addition, we find H$_2$O absorption features between 6.0-6.3 $\mu$m and 6.7-7.0 $\mu$m previously identified by \cite{Sloan2015}. In the 6.0 $\mu$m-band, we visually identify absorption lines that will be analyzed separately.

{For completeness, we note additional weak structures that may be consistent with SiO bands at $\sim$ 8~$\mu$m \citep{Sloan2015}. Furthermore, absorption features in the $\sim$ 15~$\mu$m region could be related to CO$_2$ \citep{Justtanont2004J} with potential OH band contributions in the 14-17~$\mu$m range \citep{Sloan2015}. Due to the presence of silicate dust, a H$_2$O-ice libration band around $\sim$ 12~$\mu$m can suffer from increased confusion \citep{Robinson2014}. In general, a contamination of ices in the dereddened spectrum of IRS 3 can not be excluded \citep{Moultaka2005, moultaka2015}. However, a dedicated analysis of these weaker bands is beyond the scope of this work and will be addressed in a forthcoming work. In this work, we focus on the most dominant spectral features of the AGB star IRS 3 around 6.1 $\mu$m, 9.7 $\mu$m, and 18.5 $\mu$m.}

\subsection{Optical depth}

\begin{figure}[ht!]
	\centering
	\includegraphics[width=0.5\textwidth]{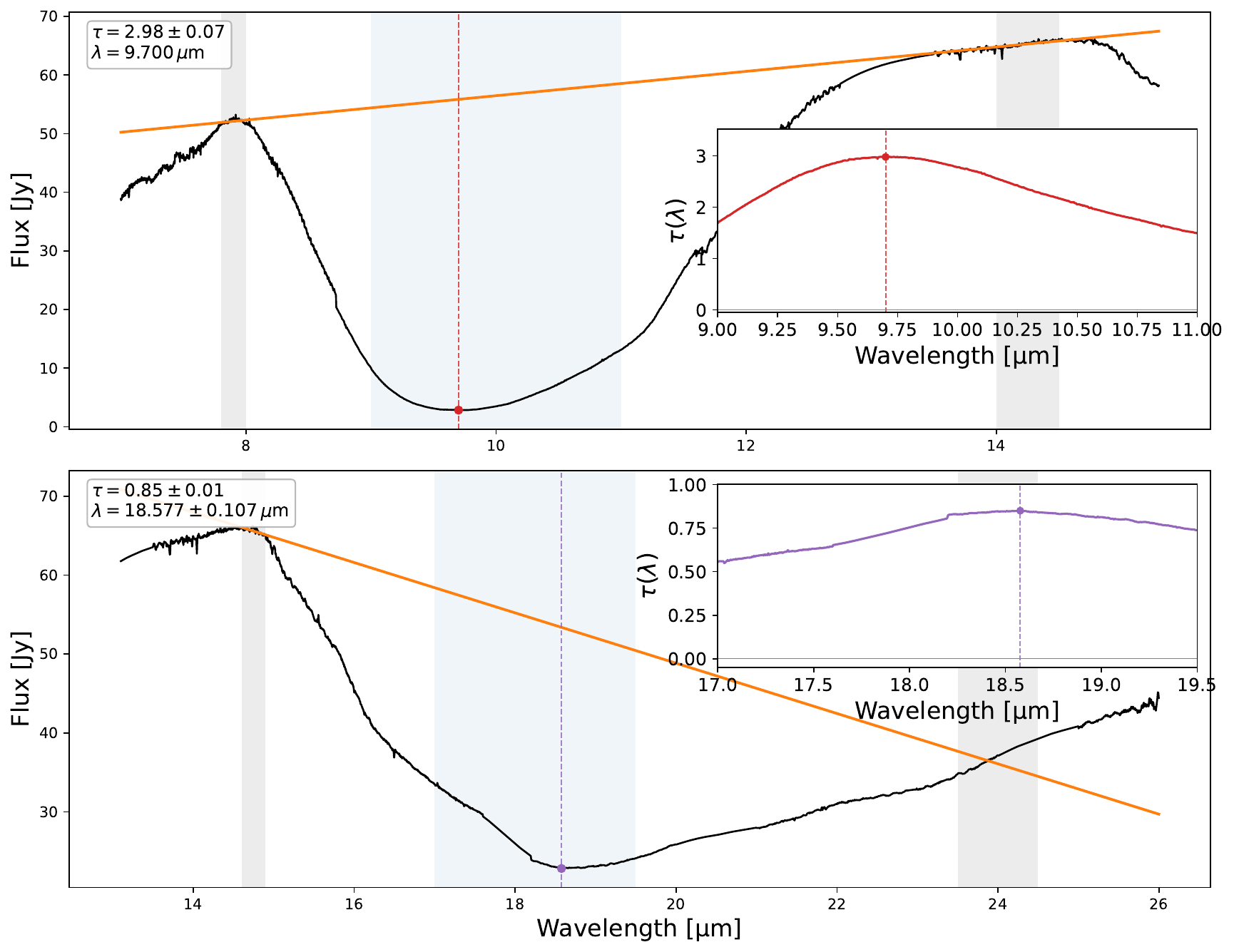}
	\caption{Optical depth estimates of the $\rm 9.7\,\mu m$ and $\rm 18.5\,\mu m$ absorption feature. The upper plot shows the $\rm 9.7\,\mu m$-, the lower the $\rm 18.5\,\mu m$-silicate absorption. Both plots exhibit an inlet that displays the related fit of the optical depth. The anchor points of the fit are shaded in gray, whereas the spectral range of the peak of the optical depth is indicated in light blue.}
\label{fig:tau_measurement}
\end{figure}

The optical depth is a crucial diagnostic parameter for further distinguishing between the dominant chemistry of an AGB star. For IRS~3, the optical depth $\tau$ of the dereddened spectrum is estimated by fixing a linear function left and right of the respective silicate absorption feature. For the Si-O silicate stretching feature at 9.7 $\mu$m, the lower fix point is at $\approx$ 8 $\mu$m and the upper fix point is at $\approx$ 14 $\mu$m. The O-Si-O silicate bending feature at 18 $\mu$m is enclosed by fixing the linear function to $\approx$ 14.5 $\mu$m and $\approx$ 24 $\mu$m (Fig. \ref{fig:tau_measurement}). Please note that the $\approx$ 24 $\mu$m fixing point is chosen such that the O-Si-O silicate bending feature at 18 $\mu$m is symmetrically enclosed. By varying the upper anchor points between 23 $\mu$m and 25 $\mu$m, we find that the estimated optical depth of the O-Si-O silicate feature is robust against the choice of fixing points in this wavelength range.

Starting from the fit of the linear function, we gain a baseline from which the ratio between observed flux F and continuum flux F$_{\rm cont}$ is measured with
\begin{equation}
    \rm \tau(\lambda)\,=\,-ln(F(\lambda)/F_{cont}(\lambda))
\end{equation}
where $\tau$ represents the optical depth. The exact process and the $\tau$ measurement is demonstrated in Fig. \ref{fig:tau_measurement}.

For the optical depth of the $\rm 9.7\,\mu m$ absorption feature, we measure $\tau_{9.7}\,=\,2.98\pm0.07$. The O-Si-O bending mode at $\rm 18.5\mu m$ yields an optical depth of $\tau_{18.5}\,=\,0.85\pm0.01$. The resulting ratio between $\tau_{9.7}$ and $\tau_{18.5}$ is $3.5\pm0.1$. Both features are consistent with amorphous silicate dust \citep{Suh2016}. For the uncertainty, we implemented Gaussian noise around the anchor/fixed points and estimated the 1$\sigma$ uncertainty for the corresponding $\tau$ value.
Given that \cite{Pott2008} measured a combination of foreground and source optical depth, resulting in an effective estimate of about $\tau_{9.7}\,\approx\,3.5$, we find a consistent value of $\tau_{9.7}\,=\,2.98\pm0.07$. Due to difficulties in calibrating ground-based Q-band data, the models of \cite{Pott2008} treated the flux density around 20 $\mu$m with caution. Our MIRI MRS data unambiguously confirm the presence of the O-Si-O bending mode around $\rm 18.5\,\mu m$.

\subsection{Model results}

\begin{figure}[htbp!]
	\centering
	\includegraphics[width=.5\textwidth]{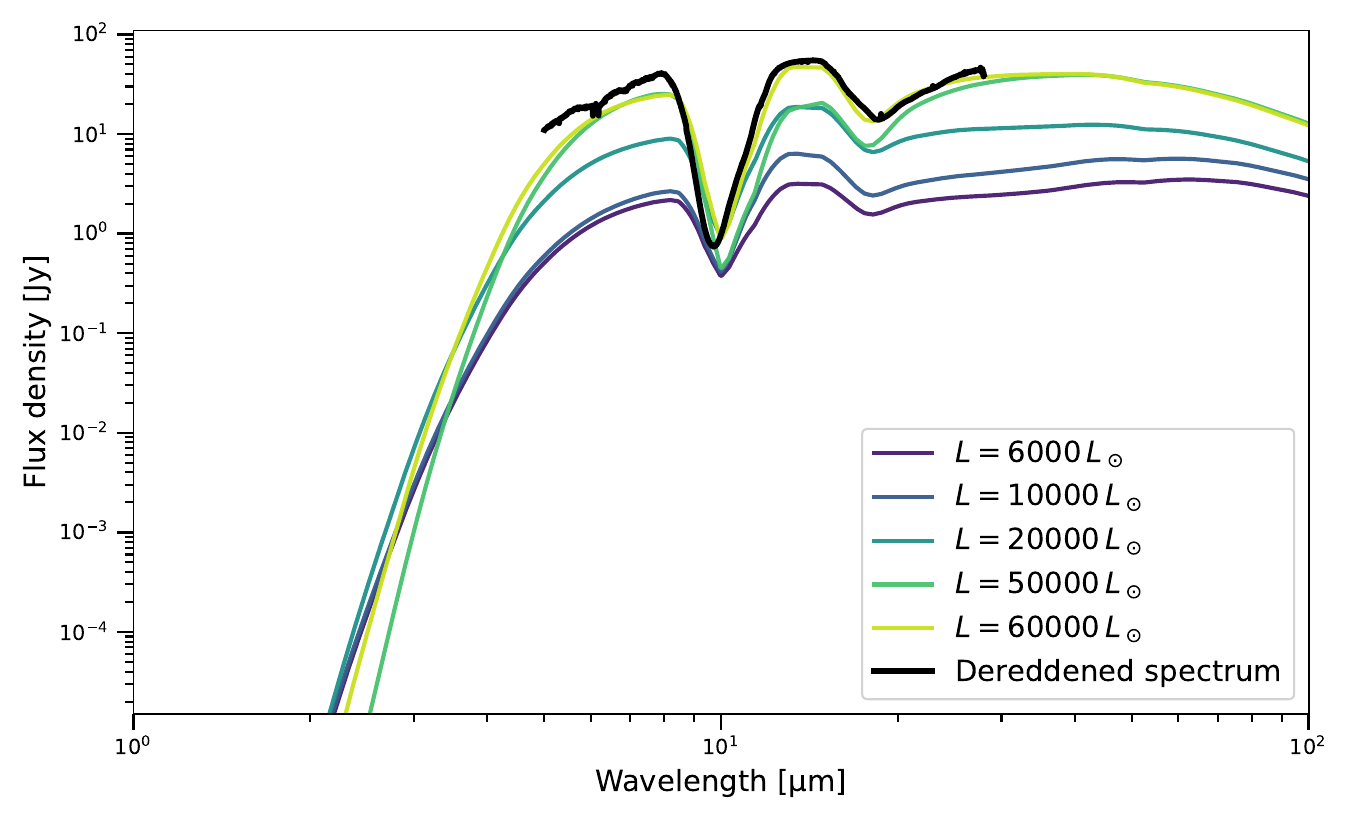}
	\caption{Comparison of the dereddened spectrum and the results of the radiative transfer modeling. For all models, we use the same parameter range as listed in Table \ref{tab:irs3_parameter_ranges}. Every SED represents a different luminosity between 6000 L$_{\odot}$ and 60000 L$_{\odot}$ and covers a spectral range of 2-100 $\mu$m.}
\label{fig:hyperion_best_rmse}
\end{figure}

In this section, we present the results of the SED modeling using Hyperion. Our main objective is to reproduce the observed spectrum and constrain the dust composition and envelope structure of IRS~3. For this, we used a parameter grid comprising 10$^5$ individual models. As mentioned in Sec. \ref{sec:data}, these models suffered from a premature setup, i.e., only included a blackbody with one shell. These settings largely underestimated the complex morphology since IRS~3 is a known bow shock source in the {inner parsec} interacting with the ISM. From the final model setup, {we perform a luminosity scan producing multiple solutions that are evaluated using} the minimized root mean square error (RMSE). {Ultimately, we compare} the observed and {dereddened spectrum and show the results} in Fig. \ref{fig:hyperion_best_rmse}.
{For the presented model results, we require} an additional cool third shell with a higher density exponent p following
\begin{equation}
    \rm \rho(r) \,\propto\,r^{-p} 
    \label{eq:density_radius}
\end{equation}
where r denotes the radius r and $\rho$ the density. {A} two-shell model {would overestimate} the flux density longward of the 18 $\mu$m absorption feature. We notice {four} interesting byproducts of the Hyperion modeling:
\begin{itemize}
\item 
{The luminosity scan between 6000 L$_{\odot}$ and 60000 L$_{\odot}$ shows that increasing the stellar luminosity improves the agreement with the observed SED at wavelengths longward of $\approx\,9\,\mu$m. Especially the dereddened flux density between 10-28 $\mu$m is sufficiently represented by the 60000 L$_{\odot}$ model setup. However, we notice no substantial difference between the 50000 L$_{\odot}$ and 60000 L$_{\odot}$ models, suggesting a luminosity saturation. This indicates that the emission is dominated by an additional component not captured by the spherically symmetric dust radiative transfer model. Possible contributors include a bow-shock interaction with the surrounding medium or anisotropic inner clumpy structures. We therefore refrain from interpreting the short-wavelength excess between 5-9 $\mu$m in terms of stellar luminosity alone.}
\item 
{As listed in Table \ref{tab:irs3_parameter_ranges}, shell 1 shows a density exponent of 0.1-0.2, which translates to an almost constant density over the modeled region. Since Eq. \ref{eq:density_radius} is only valid for constant outflow velocities v$_{\rm out}(r)$, the more general description of the shell density}
\begin{equation}
    \rm \rm \rho(r) \,\propto\,\frac{\dot{M}}{4\pi r^{p} v_{out}(r)}  
\end{equation}
{allows for variations of v$_{\rm out}(r)$ under the boundary condition of a constant density exponent p. One possible interpretation of the nature of shell 1 would be swept up material or a zone with an enhanced density. {This} could be a transition zone between the inner region and shell 2.}
\item
{A second region of interest is shell 2.} Although IRS~3 is a known bow shock source, this component is not addressed directly in the radiative transfer model. The characteristic radius of shell 2 is between shell 1 and shell 3, but shows a slightly increased density p at about {2300 AU} (Table \ref{tab:irs3_parameter_ranges}). Notably, the characteristic radius of shell 2 {partially} overlaps with the observed stand-off distance of the bow shock, which is located at about {3000 AU}. This suggests that part of the emission may be associated with the not modeled bow-shock region. {Since bow shocks represent compressed material due to the ram pressure, shell 2 can be interpreted as a density enhanced region with $\rm \rho_0\,=\,3-10\times10^{-20}$ g cm$^{-3}$.}
\item 
{Finally, we notice a mismatch between the modeled silicate absorption feature and the dereddened spectrum at around $9.7-10\,\mu$m. While this mismatch could be explained by the overall variations between the model and the spectrum, the spectral location of the silicate absorption feature is constantly at higher wavelengths for all radiative transfer runs. \cite{Kemper2004} argued that this mismatch might be explained by porosity and different grain sizes not covered by our available dust models. Speculatively, ices along the line-of-sight that contaminate our extinction models may impact the silicate absorption feature.}
\end{itemize}
For the dust composition of the models displayed in Fig. \ref{fig:hyperion_best_rmse}, a mixture of Al$_2$O$_3$ and silicates for an O-rich late-type star are used \citep{Begemann1997, Karovicova2013, Gobrecht2016}. For the inner region, we use a mix of alumina and silicates. For the shells 1-3, only silicates are used. Consequently, the temperature gradient ranges from 1200 K for the inner regions to 80-100 K for the outer dust components.

\subsection{Spectral diagnostic}

From the HITRAN database, we downloaded H$_2$O transition lists and convolve the spectral lines with the MIRI PSF. Next, we fitted these line lists to the region around 6.0 $\mu$m that shows prominent H$_2$O absorption (Fig. \ref{fig:IRS3_spectrum}). Between 6.0-6.25 $\mu$m, we identify a prominent absorption band that matches H$_2$O transition lines. As mentioned in Sec. \ref{sec:data}, we used a Voigt profile for the transition lines in combination with Gaussian residual components to account for residuals. These Gaussian models are listed in Table \ref{tab:gaussian_components}.
\begin{table}[h!]
    \centering
    \caption{Gaussian absorption components G0-G4 to approximate residual features in the 6 $\mu$m H$_2$O band.}
    \begin{tabular}{|ccccc|}
        \hline \hline
         & Type & $\lambda_0$ [$\mu$m] & FWHM [$\mu$m] & $\tau_0$ \\
        \hline
        G0 & line  & 6.0500 & 0.0008 & 0.12 \\
        G1 & broad & 6.0600 & 0.0100 & 0.03 \\
        G2 & line  & 6.1170 & 0.0008 & 0.15 \\
        G3 & line  & 6.1860 & 0.0012 & 0.14 \\
        G4 & broad & 6.185--6.195 & 0.020--0.036 & 0.02--0.05 \\
        \hline
    \end{tabular}
    \tablefoot{While the components G0-G3 are fixed, G4 is determined from a parameter scan and marks the broadest feature in the model. The Gaussian residual components G0, G2, and G3 are related to the H$_2$O transition lines from the HITRAN database.}
    \label{tab:gaussian_components}
\end{table}
We also detect broad absorption features around 6.06 $\mu$m and between 6.12-6.25 $\mu$m. Although we find a narrow H$_2$O absorption at 6.19 $\mu$m, the morphology of this feature is ambiguous and have been fitted with the Gaussian residual components listed in Table \ref{tab:gaussian_components}. Quantitatively, \cite{Keane2001} and \cite{Gibb2002} argue about H$_2$O ice refractory matter impacting the 5-8 $\mu$m spectral range.
\begin{figure}[htbp!]
	\centering
	\includegraphics[width=.5\textwidth]{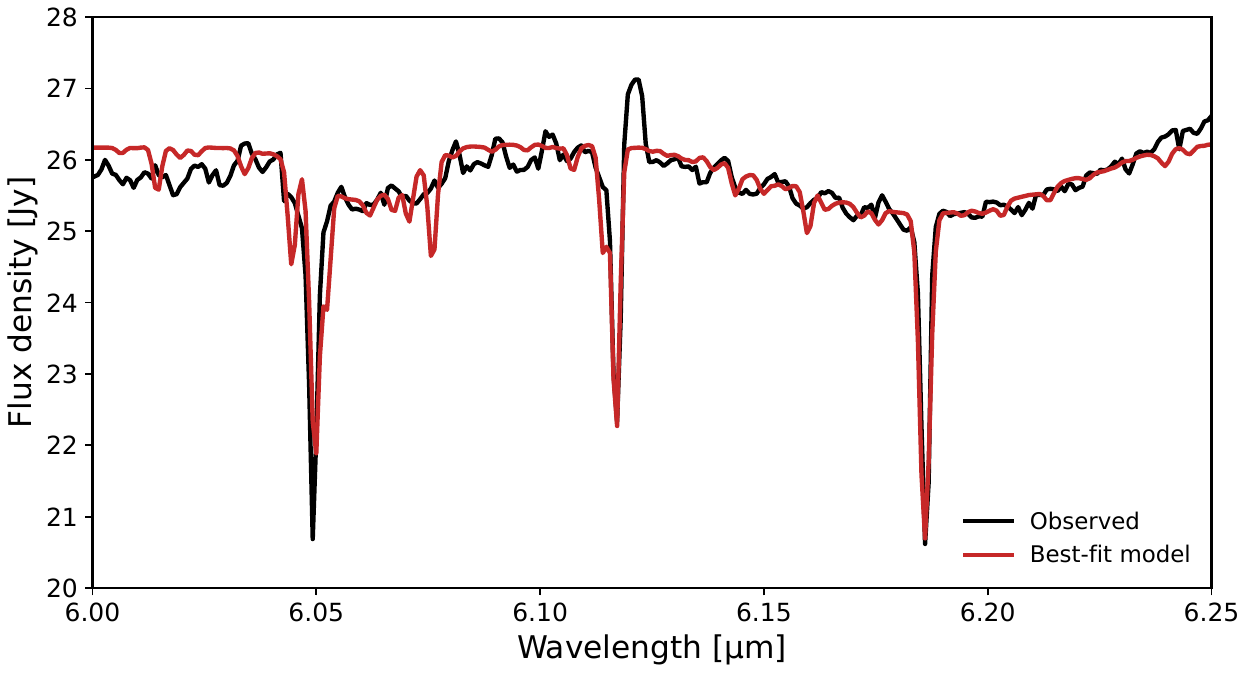}
	\caption{Synthetic spectrum based on HITRAN line list of H$_2$O overlaid on the observed spectrum of IRS~3 between 6.00-6.25 $\mu$m. The 0.1 $\mu$m wide absorption feature between 6.12-6.25 $\mu$m represents an additional water component at 6.19 $\mu$m coupled with an unidentified species.}
\label{fig:hitran}
\end{figure}
Although the exact nature of the absorption around 6.06 $\mu$m and between 6.12-6.25 $\mu$m remains unclear, a Gaussian with a characteristic width of about 0.025 $\mu$m approximates this feature (Table \ref{tab:gaussian_components}).

While the nature of the additional broad components remains ambiguous, these species could speculatively represent ice {(such as NH$_3$ or NH$^+_4$) or large molecules as outlined by \cite{Schutte2003}, \cite{Gibb2004}, \cite{Boogert2015}, and \cite{Ricca2021}}. Alternatively, the broad feature could be a residual of the extinction correction and related to foreground emission \citep{Zeegers2025}. A possible candidate for the broad feature could be foreground PAH compounds at 6.22 $\mu$m \citep{Smith2007}.

\section{Discussion} 
\label{sec:discuss_conclusion}

In this section, we discuss our finding of the MIRI MRS observations of IRS~3. We furthermore test the stellar classification by comparing the spectrum after applying all three available extinction laws. In addition, we use the bow shock of IRS~3 to derive the mass-loss rate of the AGB star.

\subsection{Stellar classification}

In this work, we have identified prominent Si-O and O-Si-O silicate stretching and bending features in the mid-infrared spectrum of IRS~3 located at $\rm 9.7\,\mu m$ and $\rm 18.5\,\mu m$, respectively. As shown in the comprehensive ISO census of \cite{Kraemer2002} and \cite{Sloan2003}, the Si-O and O-Si-O silicate features are found in AGB stars. It should be noted that at least the $\rm 9.7\,\mu m$ silicate stretching feature has been reported in Young Stellar Objects (YSO) \citep{Kessler-Silacci2005}. Historically, the stellar classification of IRS~3 underwent several revisions due to observational constraints \citep{Rieke1978, Roche1985, Krabbe1995, Viehmann2005}. For example, the ISO beam of the Galactic Center observations is too large to identify IRS~3 as a single source \citep{Kemper2004}. Although Spitzer surveys \citep{Woods2011, Jones2012} could have overcome the spatial resolution limitation of ISO, IRS~3 {was} not observed. However, \cite{Pott2008} showed the first VLTI high-angular resolution observations and classified IRS~3 as an AGB star. Furthermore, \cite{peissker2023c} showed that the flux density values between the infrared and submillimeter are not compatible with a radiative transfer model describing a YSO, strengthening the classification of IRS~3 as an AGB star. 

Interestingly, the observations presented in \cite{Pott2008} showed discrepancies in the flux density around $\sim20\,\mu$m and their SED fit, leading the authors to suggest that IRS~3 may be a carbon‑rich AGB star. However, the presence of the O-Si-O bending mode around $\rm 18.5\,\mu m$ presented in this work questions the classification of IRS~3 as a carbon‑rich AGB star.

{Furthermore, we find clear signatures of water between 6.00-6.25 $\rm \mu$m in the derreddened spectrum of IRS 3. As listed in Table \ref{tab:irs3_parameter_ranges} and mentioned in Sec. \ref{sec:results}, shell 1 shows an almost constant density. Speculatively, the water could be located in this shell at a distance of about 900 AU to the star \citep{Justtanont2006, Cherchneff2011, Gobrecht2016}.}
Although the presence of water shown in Fig. \ref{fig:hitran} in the envelope of IRS 3 is not a discriminant between carbon‑ and oxygen‑rich chemistry \citep{Decin2010, Neufeld2011, Lombaert2016}, the ratio between the Si-O silicate stretching feature at 9.7$\mu$m and the O-Si-O bending mode is a clear diagnostic of an O-rich AGB star with a silicate dust dominated chemistry. Typically, C-rich AGB stars show silicon carbide absorption features at 11.3$\,\mu$m in combination with a smooth spectrum up until 40.0$\,\mu$m \citep{Speck2009}. 

Due to carbon-based chemistry, C-rich stars lack an analog to the observed Si-O and O-Si-O bending mode. Therefore, C-rich AGB stars cannot reproduce the $\tau_{9.7}$ to $\tau_{18.5}$ ratio of $3.5\pm0.1$ \citep{Jones2012,Sloan2014}. Consequently, our analysis demonstrates that the spectral shape of IRS~3 and its deep silicate absorption can naturally only be reproduced by an oxygen-rich dust model \citep{Ossenkopf1992}. The observed spectral shape and radiative transfer models support the classification of IRS~3 as an O‑rich AGB star. Despite possible evolutionary differences, both O‑rich and C‑rich AGB stars undergo strong radial pulsations, which drive episodic mass loss. However, these pulsations occur on timescales incompatible with our preferred multi-shell model setup \citep{Ohnaka2017, Hoefner2018}. While we cannot exclude the presence of additional shells below the detection or resolution limit, the origin of the modeled envelope remains ambiguous. One possible explanation would be time-variable mass‑loss outflows, without invoking a transition to carbon‑rich chemistry. Another idea is a potential tidal interaction between IRS~3 and Sgr~A* as discussed in \cite{Bhat2022}.

In the process of finishing this study and after the modeling, we recognize the publication by \cite{Yusef-Zadeh2017-IRS3} that already reported a multi-shell structure of IRS~3. The authors of Yusef-Zadeh et al. report observations of two shells associated with IRS~3 utilizing ALMA. This strengthens our analysis and interpretation of the radiative transfer modeling that IRS~3 can be described with multiple shells. In comparison with O-rich AGB stars outside of the Galactic Center, GX Monocerotis in the unicorn star sign constellation shows a complex multi-shell setup \citep{Karovicova2013, Randall2020}.

Considering the temperature gradient and the different regions modeled, it is implied that episodic outflows are the origin of dust condensation in the extended atmosphere. These time-variable outflows might result in the formation of an envelope with a shell-like dust distribution {potentially including} Al$_2$O$_3$ and amorphous silicates.
{We note that Al$_2$O$_3$ in oxygen-rich outflows is expected to condense closer to the star at high temperatures, potentially acting as an early condensate or seed population for subsequent dust growth. With increasing mass loss rates, the spectral signature of alumina could be masked by the much stronger silicate bands if silicates form mantles or rims on early high-temperature condensates \cite{Maldoni2005}. This is consistent with the observational trend that alumina signatures are detected in oxygen-rich AGB stars with thinner dust envelopes and lower mass-loss rates \cite{Sloan1998}. In contrast, higher mass-loss rates tend to produce spectra that are dominated by silicate features.}

While the O–Si–O bending mode at $\sim18.5\,\mu$m is an intrinsic vibrational feature of silicate dust, its observed optical depth directly traces the cumulative silicate column density produced by outflow‑driven dust formation.

\subsection{Robustness of the stellar classification}

The line of sight extinction towards the {inner parsec} has been subject to various studies. For about four decades, studies have concluded an overabundance of silicates in the ISM. This finding might not be surprising since observations suggest an increasing metallicity towards the Galactic Center \citep{Do2018, Gallego-Cano2026}. \cite{Kemper2004} argue, that the increased metallicity is the reason for the porosity of silicate grains reflected in increased 9.7 $\mu$m resonances. {These} statements {may} hold for {many substructures of} the {inner parsec}, {although} \cite{Fellenberg2025} argues {for} local extinction variations that have been reported by \cite{Peissker2020c}. Inspecting the extinction map derived from \cite{Scoville2003} draws an ambiguous picture of the foreground emission of the {inner parsec} \citep{Dinh2024}.

\begin{figure}[htbp!]
	\centering
	\includegraphics[width=.5\textwidth]{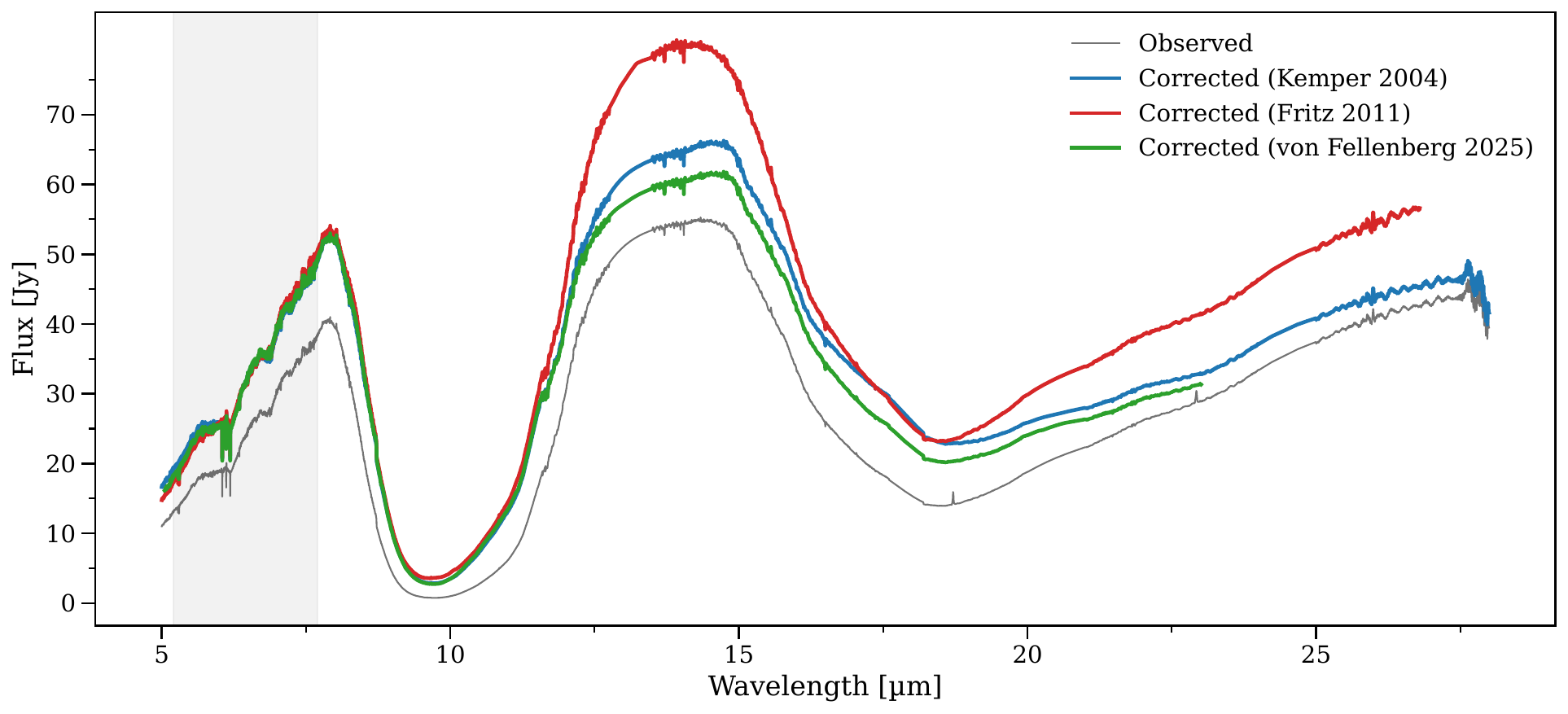}
	\caption{Resulting dereddened spectrum of IRS~3 using three different extinction laws. For all three final spectra, the classification of IRS~3 as an O-rich AGB star holds.}
\label{fig:stellar_robustness}
\end{figure}

Therefore, we apply the extinction laws of \cite{Fritz2011} and \cite{Fellenberg2025} on the observed spectrum of IRS~3 to inspect the robustness of our stellar classification. Figure \ref{fig:stellar_robustness} shows the comparison between the observed spectrum and the dereddened SEDs using all three extinction laws.
While there are deviations between 12.5-15.0 $\mu$m and 20.0-25.0 $\mu$m, the overall morphology is robust against the choice of foreground correction. All three extinction laws result in the same overall shape of the IRS~3 spectrum, strengthening our stellar classification as an O-rich AGB star. Furthermore, we notice the impact of the ISM silicate grains by the reduced width of the 9.7 $\mu$m absorption feature.

As argued before, the majority of the silicate absorption is intrinsic and produced by the envelope of IRS~3.

\subsection{Multiple shells in the envelope of IRS~3}
\label{sec:multiple_shells}

As mentioned before, we realized the publication of \cite{Yusef-Zadeh2017-IRS3} describing independently two shells of IRS~3 observed with ALMA, including signatures of a gravitational interaction with Sgr~A*. This observation led the authors of Yusef-Zadeh et al. to suggest the presence of ``fingers'', indicating large scale elongated filaments that get gravitationally attracted by Sgr~A*. Due to the resolution, we can neither confirm nor disprove the presence of these ``fingers'' because of the different wavelength regimes. While ALMA traces cold dust mostly associated with envelope regions at a distance of $\rm \rm 16\times10^4$ AU from the central star, MIRI MRS mid-infrared observations target warmer regimes at distances up to $\rm 10^4$ AU. Since our Hyperion model prefers a three shell setup compared to the mid-infrared MIRI MRS observations, two additional shells as reported by \cite{Yusef-Zadeh2017-IRS3} at an even larger distance of $\rm 16\times10^4$ AU seems reasonable. The implications of this polychromatic description of the envelope of IRS~3 are twofold.

First, the observations of two cold dust shells are consistent with our radiative transfer model described in Sec. \ref{sec:data}. Second, two additional shells that might tidally interact with Sgr~A* provide an insight into the shielding mechanisms of the envelope of IRS~3. 

If we adopt a typical AGB wind velocity of $\rm v_{\rm w} \approx 15\,{\rm km\,s^{-1}}$ \citep[][]{Vassiliadis1993}, we can provide a rough estimate about the expansion times of the individual shells by further including the characteristic radii listed in Table \ref{tab:irs3_parameter_ranges}. Using the shells located at {949 AU, 2214 AU, 6325 AU,} and $10^4$ AU as the outer radius observed with MIRI MRS yields an approximate expansion time of {300 yr, 700 yr, 2000 yr,} and 3200 yr, respectively. The outer envelope of IRS~3 with a characteristic radius of $\rm \rm 16\times10^4$ AU mentioned in \cite{Yusef-Zadeh2017-IRS3} yields an expansion time of 5000 yr. These timescales are compatible with the interpretation by Yusef-Zadeh et al. that the outermost material in the envelope of IRS~3 may tidally interact with Sgr A*.

In this picture, MIRI MRS observations trace the recent mass-loss history of IRS~3 compared to older ejecta visible with ALMA. The shell-like distribution scenario might be compatible with time-variable outflows or mechanisms driven by a putative interaction with Sgr~A* \citep{Yusef-Zadeh2017-IRS3, Bhat2022}. However, a speculative binary interaction may also contribute to complex AGB envelope morphologies, as discussed by \cite{Decin2020}.

{We note that $\rm v_{\rm w} \approx 15\,{\rm km\,s^{-1}}$ refers to a characteristic wind velocity used for our order-of-magnitude estimates. We do not exclude local deviations of $\rm v_{\rm w}$ that may occur in the wind–ISM interaction (bow-shock) region in shell 2. Follow-up studies with designated hydrodynamical simulations may resolve local wind velocity regions in the envelope of IRS 3.}

\subsection{Mass-loss rate of IRS~3 and ISM density}

\begin{figure}[htbp!]
	\centering
	\includegraphics[width=.5\textwidth]{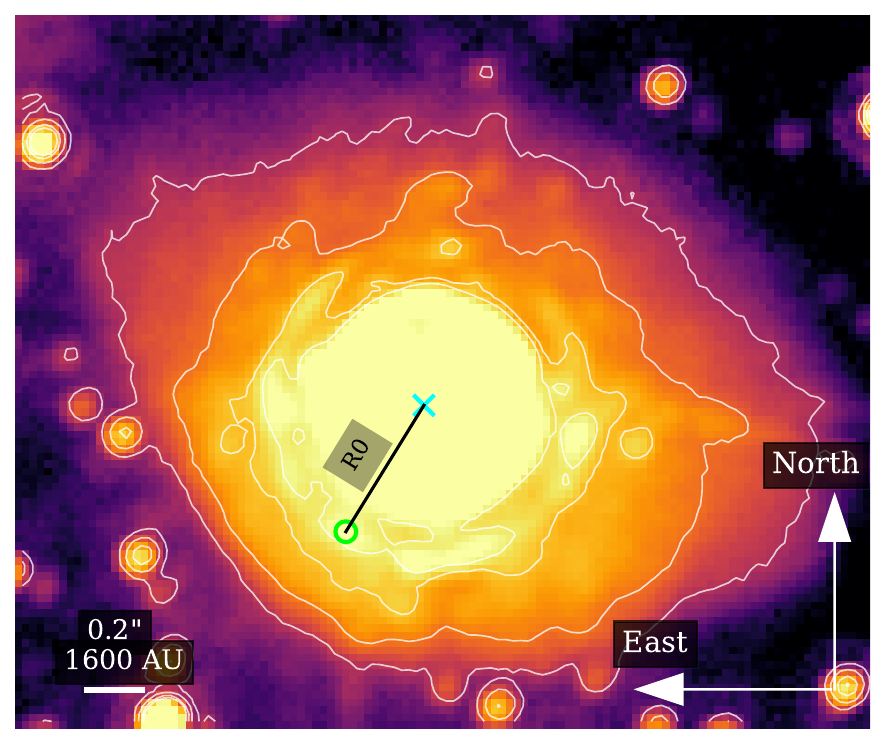}
	\caption{Projected stand-off distance estimate of the bow shock of IRS 3. The stand-off distance is a measure of the ambient ISM and the ram pressure of the star. Here, the green x marks the position of IRS 3 and the green circle the apex of the bow shock.}
\label{fig:standoff_distance}
\end{figure}

First mentioned in \cite{Viehmann2005}, the AGB star IRS~3 shows a prominent bow shock, especially visible in the L- and M-band (Fig. \ref{fig:finding_chart}). As suggested by \cite{Zajacek2020}, this bow shock may be produced by the interaction with the ISM \citep{Wilkin1996, Wilkin2000, Christie2016}. Intriguingly, this bow shock can be used to estimate the mass loss rate ($\dot{M}$) of the AGB star. As mentioned before and used in the Hyperion model, the total extent of the dusty envelope is about $10^4$~AU. {By simply measuring the distance between the bow shock and the star (Fig. \ref{fig:standoff_distance}), we find the projected stand-off distance of $R_0 \approx 4316$~AU.} 
{At the stand-off distance, the ram pressure of the AGB wind balances the ram pressure of the ambient ISM \citep{Wilkin1996, Zajacek2020}. Since the concentric rings visible in Fig. \ref{fig:standoff_distance} are produced by the diffraction limit of the telescope, our estimated projected stand-off distance serves as an upper limit. Based on \cite{vanLoon2005}, we use the empirical relation}
\begin{equation}
     \rm log(\dot{M}) = -5.65 + 1.05 log\left(\frac{L}{10 000 {L}_{\odot}}\right) -6.3log\left(\frac{T_{eff}}{3500 K}\right)
     \label{eq:van-loon}
\end{equation}
{where the estimated luminosity of 60000 L$_{\odot}$ and effective temperature of 2800 K yields a mass loss rate of $\rm \approx 6 \times 10^{-5} \, {M_{\odot}}{yr^{-1}}$. Although there might be updates to the mass-loss rate estimated with eq. \ref{eq:van-loon}, the order-of-magnitude will not differ significantly \citep{Justtanont2017, Hoefner2018}.} 

The pressure equilibrium is given by $\rm \rho_{\rm w} v_{\rm w}^2\,=\,\rho_{\rm ISM} v_{\star}^2$, where $v_{\rm w}$ is the terminal wind velocity. Furthermore, $v_{\star}$ describes the 3D space velocity of the star relative to the ISM, and $\rho_{\rm ISM} = \mu m_{\rm H} n_{\rm H}$ is the ambient mass density. Using the continuity equation for a spherical outflow, $\dot{M} = 4\pi R_0^2 \rho_{\rm w} v_{\rm w}$, the mass-loss rate can be expressed as:
\begin{equation}
    \dot{M} = 4\pi R_0^2 \, \mu m_{\rm H} n_{\rm H} \frac{v_{\star}^2}{v_{\rm w}}.
    \label{eq:mass_loss}
\end{equation}
As in Sec. \ref{sec:multiple_shells}, we assume a typical AGB wind velocity of $v_{\rm w} \approx 15\,{\rm km\,s^{-1}}$ \citep[][]{Vassiliadis1993} and a velocity of $\rm v_{\star} \approx 100\,{\rm km\,s^{-1}}$ adopted from \cite{Pott2008}. {Finally, we use IRS~3 itself as a test particle and invert Eq.~(\ref{eq:mass_loss}) to infer the local ambient number density required at the bow-shock apex to reproduce the measured stand-off distance R$_0\,=\,4316$ AU as shown in Fig. \ref{fig:standoff_distance}. Therefore, we use the above parameters to estimate an effective density at the bow-shock apex n$\rm _H$ that potentially explains the dimensions of the envelope of IRS 3:}
\begin{equation}
\begin{split}
n_{\rm H} \approx\; & 4.5\times 10^{2}\,\mathrm{cm^{-3}}
\left(\frac{\dot{M}}{6\times 10^{-5}\,M_{\odot}\,\mathrm{yr^{-1}}}\right)
\left(\frac{R_0}{4316\,\mathrm{AU}}\right)^{-2} \\
& \times
\left(\frac{v_{\star}}{10^{2}\,\mathrm{km\,s^{-1}}}\right)^{-2}
\left(\frac{v_{\rm w}}{15\,\mathrm{km\,s^{-1}}}\right)
\left(\frac{\mu}{1.4}\right)^{-1}.
\end{split}
\label{eq:nH_scaling_new}
\end{equation}
{If we assume fully ionized gas, we can set n$\rm _H\,\sim\,$n$\rm _e$. In this case, our estimate of n$\rm _H\,\sim\,10^2 cm^{-3}$ is almost two order-of-magnitudes lower compared to other works of the same region that estimate the electron density in ionized streamers \citep{Zhao2010}. Using the statistical uncertainty from \cite{Whisnant2025} for measuring the stand-off distance of $\pm\,40\%$, the resulting number density is in the range of n$\rm _H\,\approx\,(2.3\,\times\,10^2-1.2\,\times\,10^3)\, cm^{-3}$, numerically close to the models discussed in \cite{Yusef-Zadeh2017-IRS3}.}

{The different order of magnitudes suggest that the number density in the {inner parsec} is inhomogeneous, allowing for bow shocks as observed for IRS 3. For completeness, we adopt n$\rm _H\,=\,10^4 cm^{-3}$ from \cite{Zhao2010} for dense regions, which yields an unrealistically high mass loss rate of $\rm \approx 3 \times 10^{-3} \, {M_{\odot}}{yr^{-1}}$ \citep{Hoefner2018}. This example implies that stellar envelopes undergo a significant evolution due to the inhomogeneous density in the {inner parsec} as observed for other bow shock sources \cite{muzic2010, peissker2021, Ciurlo2023, peissker2023b}.}

{We further find that a mass-loss rate $\dot{M}\approx 6\times10^{-5}\,M_\odot\,{\rm yr^{-1}}$ for IRS~3} is characteristic of the superwind phase of an O-rich AGB star \citep{Hoefner2018}. 
{We note that \cite{Pott2008} and \cite{Yusef-Zadeh2017-IRS3} adopted AGB wind velocities $v_{\rm w}\approx 30\,{\rm km\,s^{-1}}$ and $\approx 20\,{\rm km\,s^{-1}}$, respectively. Interestingly, both works report a similar mass-loss rate of $\dot{M}\sim 6\times10^{-5}\,M_\odot\,{\rm yr^{-1}}$, consistent with our estimate using Eq. \ref{eq:mass_loss}. Due to different diagnostics and assumptions in all three works, a strict connection between the mass-loss rate and $v_{\rm w}$ is not necessarily applicable. In our bow-shock framework with the measured stand-off distance (Fig. \ref{fig:standoff_distance}), the ambient density scales with the wind momentum flux, $n_{\rm H}\propto \dot{M}\,v_{\rm w}$. Consequently, adopting $v_{\rm w}=20$ or $30\,{\rm km\,s^{-1}}$ would increase the inferred $n_{\rm H}$ by factors of 1.33 or 2, respectively. In contrast, keeping $n_{\rm H}$ fixed would imply $\dot{M}\propto v_{\rm w}^{-1}$.}

These calculations suggest that the observed envelope of IRS~3 shows signatures of significant and time-variable mass loss that enriches the dust reservoir in the vicinity of Sgr~A*. It also suggests that only AGB stars with a sufficient mass loss rate can be observed in the {inner parsec}. Since IRS~3 is the {most prominent} AGB source among millions of stars in one of the oldest stellar habitats in the Milky Way, this finding offers an explanation for the missing red giants. We note that the envelope stripping close to Sgr~A* may not be counterbalanced by the mass loss rate \citep{Zajacek2020}.

\subsection{Stellar tracks}

{Motivated by the radiative transfer analysis, we use the luminosity L and effective temperature of IRS 3 to inspect its the possible age and mass. We can safely assume, that the star must have formed at a sufficient distance to Sgr~A*. Consequently, a possible birthplace is located outside the {inner parsec} in the Nuclear Stellar Cluster. Migration timescales for stellar mass objects are in the range of a few parsecs \citep{Morris1993}. This suggests a distance of a potential birthplace for IRS of less than 5 parsecs. Recently, \cite{Gallego-Cano2026} investigated parts of the Nuclear Stellar cluster and found a mean super-solar metallicity of $\rm [M/H]\approx +0.35$ \citep{Feldmeier-Krause2020, schoedel2020, Nogueras-Lara2022, Feldmeier-Krause2025} and multiple stellar populations distinguished by their age.}

\begin{figure}[htbp!]
	\centering
	\includegraphics[width=.5\textwidth]{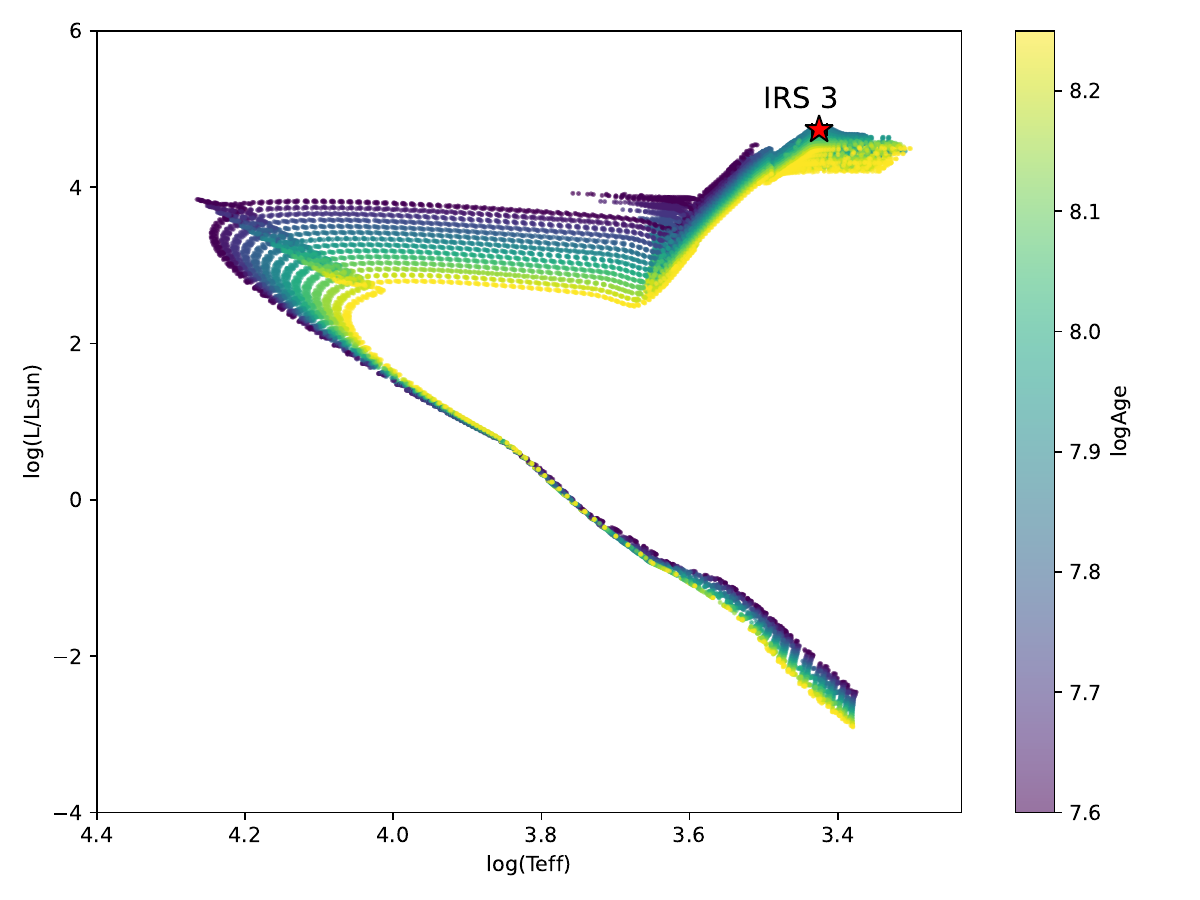}
	\caption{Hertzsprung-Russel-Diagram with isochrones showing the main-sequence (central diagonal line) and post-main-sequence (upper horizontal line) stellar tracks for low, intermediate, and massive stars with a super-solar metallicity of $\approx$+0.35. The cold and luminous AGB star IRS 3 is located in the upper right corner of the image and indicated with a red star.}
\label{fig:stellar_traccks}
\end{figure}

{For the stellar tracks, we use PARSEC-COLIBRI\footnote{\url{https://stev.oapd.inaf.it/cgi-bin/cmd}} and isochrones provided by \cite{Bressan2012, Marigo2013, Chen2014, Tang2014, Chen2015, Rosenfield2016, Pastorelli2019, Pastorelli2020, Nguyen2025}. As input parameters, we use an effective Temperature of 2800 K, a stellar luminosity of L$\,=\,(55\pm5)\,\times\,10^3\,$L$_{\odot}$, and an assumed metallicity of +0.35 dex. With these input parameters, IRS 3 is located on the post-main-sequence section of the stellar tracks as shown in Fig. \ref{fig:stellar_traccks}.}
{Based on the location of IRS 3 in the Hertzsprung-Russel-Diagram, we find a best-fit distribution of parameters to describe the star that are listed in Table \ref{tab:irs3_top10_stats}.}
\begin{table}[h!]
\centering
\caption{Results of the analysis of the stellar tracks.}
\label{tab:irs3_top10_stats}
\begin{tabular}{lcc}
\hline
Parameter & Mean & Std \\
\hline

M$\rm _{\mathrm{H}}$
 & 0.3440
 & 0.0084 \\

Z$\rm _{\mathrm{ini}}$
 & 0.030479
 & 5.31$\times 10^{-4}$ \\

$\rm\log(\mathrm{Age/yr})$
 & 7.8600
 & 0.0211 \\

M$\rm _{\mathrm{ini}}\,[M_\odot]$
 & 6.0646
 & 0.1152 \\

M$\rm\,[M_\odot]$
 & 5.9209
 & 0.0720 \\

$\rm\log(L/L_\odot)$
 & 4.7295
 & 0.0106 \\

$\rm\log(T_{\mathrm{eff}}/\mathrm{K})$
 & 3.43772
 & 0.00184 \\

$\rm\log g$
 & $-0.8161$
 & 0.0173 \\

\hline
\end{tabular}
\tablefoot{The uncertainties represent the 3$\sigma$ standard deviation of possible stellar parameters to describe IRS 3.}
\end{table}
While we find a good agreement for our input parameters, the analysis results in a stellar mass of $\rm \approx 6\,M_{\odot}$ and an age estimate of $\rm 72.4^{+3.6}_{-3.5}\,Myr$. This places IRS 3 in the range of the comparatively young stellar population of the Nuclear Stellar Cluster as described by \cite{Gallego-Cano2026}. {Speculatively, the high-mass loss rate of IRS~3 might contribute to the enrichment of dust close to Sgr~A* which is reflected by a matching metallicity \citep{Ford2026}.}

\section{Conclusion}
\label{sec:conclusion}

In this work, we present the successful observations of the Galactic Center source IRS~3, carried out with the MIRI instrument onboard the JWST in 2025. The spectrum shows two prominent absorption features at 9.7 $\mu$m and 18.5 $\mu$m. Both features are related to the presence of amorphous silicates and its respective stretching and bending modes. With the O-rich silicates used to model the emission and the multi-shell setup, we conclude that the dust production of IRS~3 is capable of enriching the ISM close to Sgr~A*. The observed and dereddened spectrum exhibits a high optical depth $\tau$ for both silicate features with a related ratio of $3.5\pm0.1$. The detection of water in the envelope furthermore shows that fundamental molecular species such as H$_2$O can survive the harsh imprint of a supermassive black hole. 

For this work, we have formulated scientific questions that are addressed by analyzing the dereddened mid-infrared spectrum of IRS~3. In contrast to the proposed carbon-rich classification of IRS~3 by \cite{Pott2008}, we find clear signatures of oxygen-rich chemistry in the analyzed spectrum. Based on estimated timescales from the modeled and observed dusty shells of the AGB star, the MIRI MRS observations trace the recent mass-loss history. The multi-shell envelope may originate from time-variable outflows, a putative companion, or an interaction with Sgr~A*. 

In the following, we summarize our key findings of this work.
\begin{enumerate}
    \item For the first time, we observe a {continuous} mid-infrared spectrum of IRS~3
    \item The morphology of the spectrum is robust against the choice of the extinction law
    \item IRS~3 is an O-rich AGB star with a stellar mass of $\rm\approx\,6\,M_{\odot}$ {and an age of $\approx$72 Myr, potentially formed in the Nuclear Stellar Cluster} 
    \item Using radiative-transfer models with {an outflow-driven} multiple shell setup, we reconstruct the dereddened spectrum of IRS~3
    \item {With a best-fit stellar luminosity of 60000 L$_{\odot}$, we} estimate a mass-loss rate of about $\rm 6 \times 10^{-5} \, M_{\odot}\,{\rm yr^{-1}}$ with an assumed wind velocity of v$\rm _w\,=\,$15 km/s
    \item For the first time, we find clear signs of H$_2$O in the envelope of IRS~3 in the {inner parsec} of our Milky Way {that resist the harsh conditions in the vicinity of Sgr~A*}
\end{enumerate}

\begin{acknowledgements}
{We thank the referee, Kay Justtanont, for the constructive and helpful comments that improved the quality of the analysis.}
FP gratefully acknowledges the Collaborative Research Center 1601 funded by the Deutsche Forschungsgemeinschaft (DFG, German Research Foundation) – SFB 1601 [sub-project A3] - 500700252. The data were obtained from the Mikulski Archive for Space Telescopes at the Space Telescope Science Institute, which is operated by the Association of Universities for Research in Astronomy, Inc., under NASA contract NAS 5-03127 for JWST. MB acknowledges funding from the Belgian Science Policy Office (BELSPO) through the PRODEX project “JWST/MIRI Science exploitation” (C4000142239). Lara Pantoni acknowledges funding from the Belgian Science Policy Office (BELSPO) through the PRODEX project 'JWST/MIRI Science exploitation' (C4000142239). AAH and LHM acknowledge support from grant PID2021-124665NB-I00 funded by MCIN/AEI/10.13039/501100011033 and by ERDF A way of making Europe.  
\end{acknowledgements}

\bibliography{bib}{}
\bibliographystyle{aasjournal}

\begin{appendix}

\section{Data Reduction and Post processing}
\label{appendix:post_processing}

The MRS data were processed with version 1.20.2 of the JWST calibration pipeline \citep{Bushouse2025}, context jwst\_1464.pmap. 

As mentioned in Sec. \ref{sec:data}, we find saturation in several bands. However, this saturation occurred in fewer than 2 groups which is why the standard pipeline was unable to derive a slope. Because of the missing slope, the affected wavelengths show severe artifacts. Especially in parts of the trace center, the saturation resulted in unusable data where the cube building algorithm is filling in from adjacent pixels, thus contaminating the signal from the central point source. To remediate this, we applied a custom data reduction method that uses an associated background exposure to estimate the detector reset point, allowing us to recover a slope from a single group in as many pixels as possible. Specifically, each science exposure was run through the calwebb\_detector1 stage of the pipeline three times, creating the following files: 
\begin{itemize}
\item a standard rate file 
\item a second-rate file, created by splicing the first group from the corresponding background uncalibrated file into the first group of the science uncalibrated file (and shifting all remaining groups back by one), while also skipping the reset switch charge decay step.
\item a third-rate file produced by running the detector stage of the pipeline normally, but skipping the jump step (which can erroneously flag 3-group ramps as cosmic rays due to known non-linearity behavior in the MIRI detectors).
\end{itemize}
We then generated a final rate file using the standard rate file as a starting point, filling in missing values only when necessary. The absolute flux calibration of such pixels will be less reliable than those provided by the regular pipeline, but the higher uncertainty is sufficient for our analysis and a significant improvement on the spectral artifacts otherwise present.

These final rate files were used as input to the subsequent pipeline stages. They were processed in a standard manner (\citep{PipelineNotebooks2025}), applying a master background subtraction with the dedicated background, and using the residual fringe correction at the time of extracting the spectra. 

Once the data was fully reduced and the IRS~3 spectrum extracted, we noticed partial discontinuities in the overlap region between channels with zero flux density. To correct for these discontinuities, we apply a polynomial of fourth order to the spectral region between 12.75-13.50 $\mu$m. We replace the observed flux with the polynomial. While this procedure removes narrow spectral information, it does not affect the broad absorption features discussed in this work. In addition to the discontinuity at 13.00 $\mu$m, spectral features at 15.5 $\mu$m, 18 $\mu$m, 21 $\mu$m, and 24.5 $\mu$m are corrected with a polynomial of third order. Additionally, we remove residual spikes with a narrow width using the SciPy Python package \citep{Scipy2020}. After the correction, we noticed some emission lines that are not included in the foreground emission. As mentioned in the text, the sulfur and forbidden iron emission line at 18.7 $\mu$m and  22.9 $\mu$m are removed with a polynomial.

\section{Reddened and Dereddened spectrum of IRS~3}
\label{appendix:raw_spectrum}

\begin{figure}[htbp!]
	\centering
	\includegraphics[width=.5\textwidth]{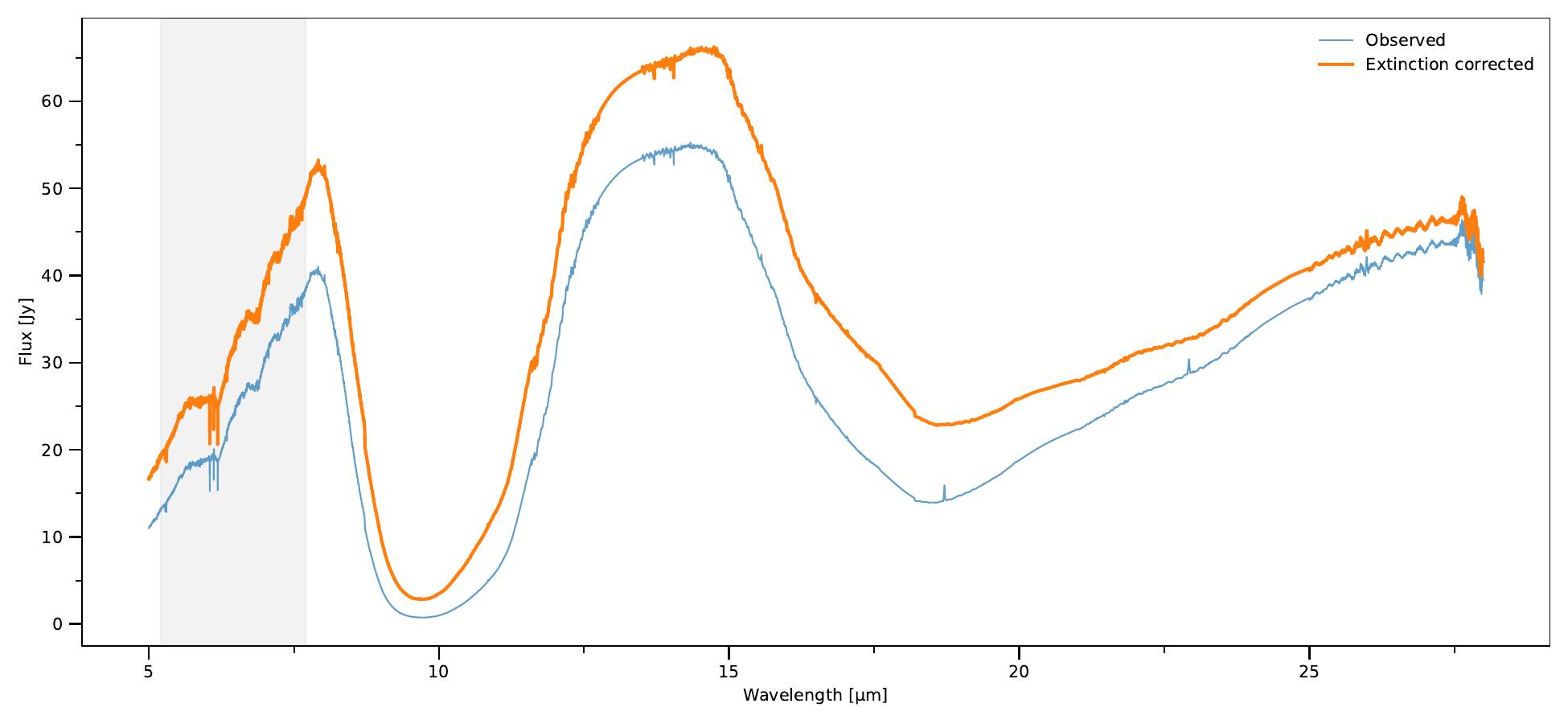}
	\caption{Comparison between the raw (blue) and extinction corrected spectrum (orange) of IRS~3. The anchor region for the extinction law applied is indicated with a gray bar.}
\label{fig:raw_spectrum}
\end{figure}

In Fig. \ref{fig:raw_spectrum}, we compare the dereddened and post-processed spectrum in orange with the reddened blue-colored spectrum. As described in Sec. \ref{sec:data}, the reddened spectrum was corrected with the extinction law by \cite{Kemper2004}. In addition to the arguments provided in Sec. \ref{sec:data}, we provide the following additional aspects to motivate our choice of the extinction law.

We applied different models and methods based on the extinction laws provided by \cite{Fritz2011} and \cite{Fellenberg2025}, as we further show in Fig. \ref{fig:stellar_robustness}. It is important to note that these models are constructed in a way that does not fully address the nature of stars at different evolutionary stages. This discrepancy can easily be explained by the intrinsic dust that is associated with the star inspected. While flux density values extracted from the models of \cite{Fritz2011} and \cite{Fellenberg2025} may produce sufficient results of individual near- and mid-infrared bands \citep{Peissker2020c, peissker2023b, Peissker2024c} and even show consistent results between different telescopes and instruments \citep{Sabha2010, Stolte2010, Peissker2024a}, we prefer the stellar-motivated extinction law provided by \cite{Kemper2004}. In the following, we discuss this decision in more detail.

The extinction law of \cite{Kemper2004} is observed with ISO, including most of the emission areas shown in Fig. \ref{fig:finding_chart}. But it is important to note that the mid-infrared emission is dominated by IRS 1W, a known Wolf-Rayet star in the Galactic Center. Therefore, the emission observed by Kemper et al. is produced by this WR star. The authors of Kemper et al. compared the observed ISO spectrum and argued that the foreground emission does not significantly differ between the Quintuplet cluster and the {inner parsec} of the Galactic Center. With this, the authors of \cite{Kemper2004} compared the spectra of Wolf-Rayet stars observed in the Quintuplet cluster \citep{Chiar2001} and the {inner parsec} of the Galactic Center \citep{Viehmann2005}. From the variations of the spectral analysis, the stellar extinction law for the GC was constructed. This extinction law is extracted from the GitHub repository\footnote{\url{ https://github.com/Sebastiano-von-Fellenberg/MIR-Extinction/tree/main}} provided by \cite{Fellenberg2025} and is normalized to the silicate feature at 9.7 $\mu$m. It is not sufficient to apply the stellar extinction law directly to the IRS 3 spectrum simply because it is normalized to a spectral feature (i.e., the silicate feature at 9.7 $\mu$m) that shows variations in depth and width due to the grain composition. Therefore, we used a region of the spectrum free from (large) spectral features as an anchor for the extinction law. From a visual inspection, we used the continuum-dominated region between 5-7.7 $\mu$m (see the gray shaded area in Fig. \ref{fig:raw_spectrum}) to align the extinction law with the observed spectrum. Naturally, we introduce an additional extinction component because the extinction law was initially normalized to 9.7 $\mu$m. This is corrected with a foreground factor of A$_{\rm fg}\,=\,0.3$ mag. Comparing the strength of the 9.7$\mu$m feature of the resulting extinction law with \cite{Fellenberg2025} yields consistent values.

Applying the derived extinction law to the observations results in the spectrum shown in Fig. \ref{fig:IRS3_spectrum}. A comparison between the extinction corrected and raw spectrum is displayed in Fig. \ref{fig:raw_spectrum}. Compared to the O-rich AGB stars in the Large Magellanic Cloud \citep{Jones2012,Jones2014}, the spectrum of IRS~3 shows a matching morphology underlying our stellar classification presented in this work. For galactic sources, the O-rich AGB star IRAS 01304+6211 shows a comparable morphology to IRS~3 \citep{Gupta2004}. Our flux density values are consistent with the VLTI observations of IRS~3 presented in \cite{Pott2008}. 

\end{appendix}

\end{document}